\documentclass[aps,prl,twocolumn,longbibliography,superscriptaddress]{revtex4-1}

\usepackage{amssymb}
\usepackage{amsbsy}
\usepackage{amsmath}
\usepackage{graphicx}
\usepackage{graphics}
\usepackage{setspace}
\usepackage{array}
\usepackage{color}
\usepackage{fontenc}
\usepackage{textcomp}
\usepackage{bm}
\usepackage{float}
\usepackage[bookmarks=false,linkcolor=blue,urlcolor=blue,colorlinks,citecolor=blue]{hyperref}

\newcommand{\vex}[1]{\bm{\mathrm{#1}}}

\newcommand{\bsub}{\begin{subequations}}
\newcommand{\esub}{\end{subequations}}

\graphicspath{{../Figures/}}

\begin{document}
\title{Higher-order Weyl Semimetals}

\author{Sayed Ali Akbar Ghorashi}\email{sghorashi@wm.edu}
\affiliation{Department of Physics, William $\&$ Mary, Williamsburg, Virginia 23187, USA}
\author{Tianhe Li}
\author{Taylor L. Hughes}
\affiliation{Department of Physics and Institute for Condensed Matter Theory,
University of Illinois at Urbana-Champaign, IL 61801, USA}

\date{\today}

\newcommand{\be}{\begin{equation}}
\newcommand{\ee}{\end{equation}}
\newcommand{\bea}{\begin{eqnarray}}
\newcommand{\eea}{\end{eqnarray}}
\newcommand{\h}{\hspace{0.30 cm}}
\newcommand{\vs}{\vspace{0.30 cm}}
\newcommand{\n}{\nonumber}
\begin{abstract}
We investigate higher-order Weyl semimetals (HOWSMs) having bulk Weyl nodes attached to both surface and hinge Fermi arcs. We identify a new type of Weyl node, that we dub a $2nd$ order Weyl node, that can be identified as a transition in momentum space in which both the Chern number and a higher order topological invariant change. As a proof of concept we use a model of stacked higher order quadrupole insulators to identify three types of WSM phases: $1st$-order, $2nd$-order, and hybrid-order. The model can also realize type-II and hybrid-tilt WSMs with various surface and hinge arcs. Moreover, we show that a measurement of charge density in the presence of magnetic flux can help identify some classes of $2nd$ order WSMs.  Remarkably, we find that coupling a $2nd$-order Weyl phase with a conventional $1st$-order one can lead to a hybrid-order topological insulator having coexisting surface cones and flat hinge arcs that are independent and not attached to each other.  Finally, we show that periodic driving can be utilized as a way for generating HOWSMs. Our results are relevant to metamaterials as well as various phases of Cd$_3$As$_2$, KMgBi, and rutile-structure PtO$_2$ that have been predicted to realize higher order Dirac semimetals.\\
\end{abstract}

\maketitle

\emph{Introduction}.-- Recently, a new family of topological crystalline phases that admit a higher-order bulk-boundary correspondence has been discovered. They have been dubbed higher-order symmetry protected topological (HOSPT) phases \cite{Benalcazar2017-1,Benalcazar2017-2,Song2017,Schindler2018-1,Langbehn2017,benalcazar2019,ghorashi2019vortex,tianhe1, Serra-Garcia2018,Peterson2018,Noh2018,Schindler2018-2,Imhof2018,xue2019acoustic,ni2019observation,PhysRevLett.124.036401,
PhysRevLett.123.196402,PhysRevLett.123.167001,PhysRevLett.123.177001,PhysRevLett.124.046801,zhang2019higherorder,PhysRevResearch.2.012018,
dubinkin2020lieb,PhysRevB.98.235102,PhysRevB.99.235132,PhysRevLett.123.216803,PhysRevLett.123.266802,PhysRevB.101.241104,
kheirkhah1,Loss2018,kheirkhah2,ZhongWang2018,FanZhang2018,yanPRL2019}, and the hallmark of \emph{nth-order} HOSPT phases is the existence of gapless states (or other observable topological features) on boundaries having co-dimension $d_c=n$. In this classification the conventional topological insulator phases are \emph{1st-order}. Moreover, the  coexistence of, for example, $d_c=n-1$ and $d_c=n$ boundary modes has been explored and is usually referred to as hybrid-order topology \cite{bultinck2019three,kooi2019hybrid,Ghorashihosc2019,Lin2017}. \\

\indent In addition to HOSPTs, very recent works have explored higher-order topological semimetals, that are often characterized by their hinge states  \cite{CAlugAru2018,Lin2017,PhysRevLett.120.026801,Wieder2020,ezawa2019second,szabo2020dirty,PhysRevLett.123.186401,PhysRevLett.124.156601,PhysRevB.101.241104}, but so far these works have primarily focused on Dirac-like semi-metal systems. In this article, we instead identify a type of higher-order Weyl semimetal (HOWSM): a class of semimetals which consists of at least a pair of what we will call $2nd$-order Weyl nodes in the bulk. In addition to attached surface Fermi arcs,  $2nd$-order Weyl nodes have Fermi hinge arcs attached too. This is distinct from the case of higher-order Dirac semimetals(HODSMs), where the possible appearance of surface states is not related to the topology of the bulk Dirac nodes, i.e., they are not required to connect to the projections of the bulk Dirac nodes on the surfaces, but are instead associated to the topology in high-symmetry planes \cite{Lin2017,Wieder2020,CAlugAru2018}.
\begin{figure}[h!]
\centering
    \includegraphics[width=\columnwidth]{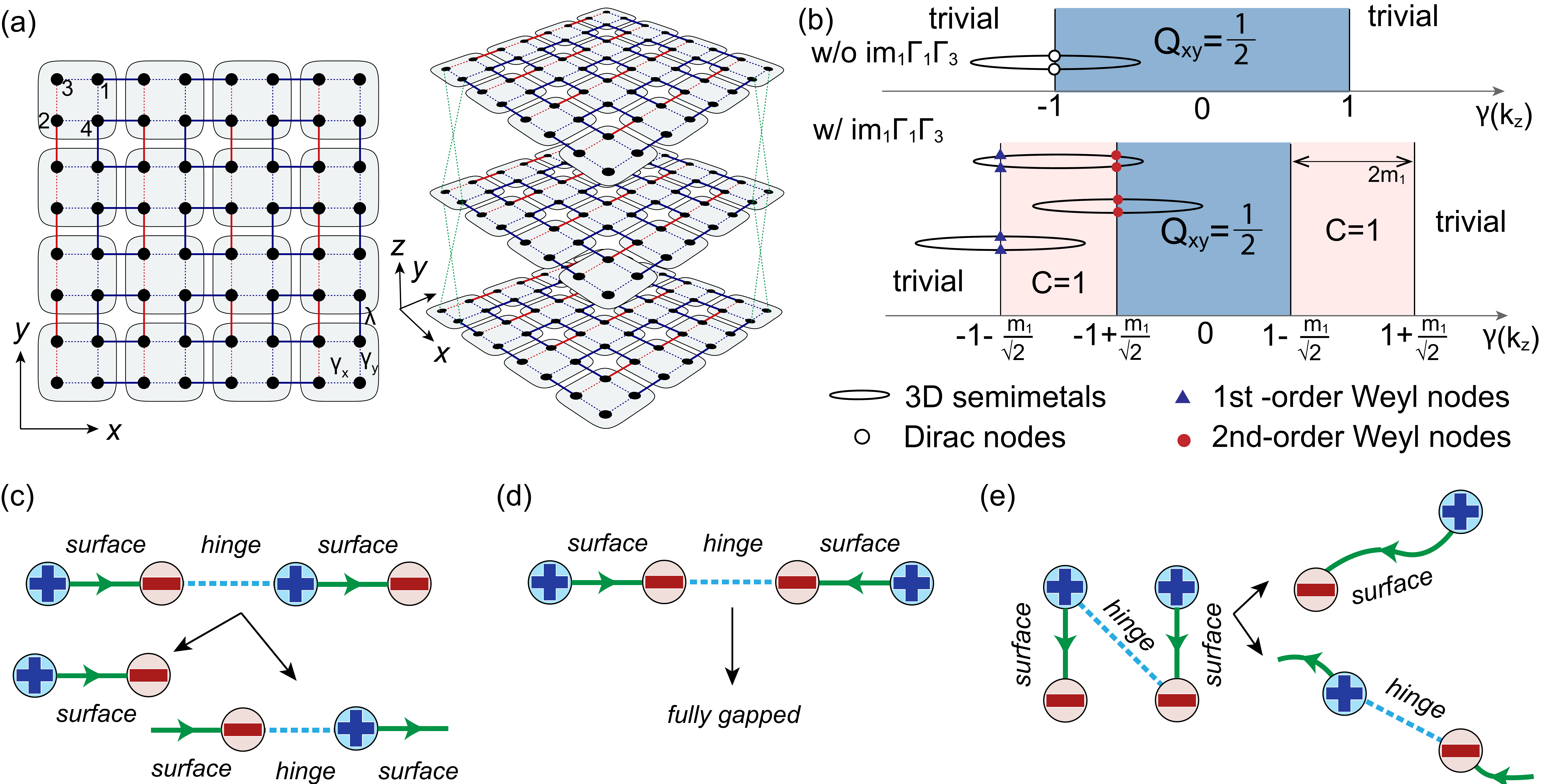}\\
    \caption{(a) The lattice configuration of the higher-order Dirac semimetal model, which is constructed by stacking quadrupole insulators (QI). (b) Phase diagram of the $2D$ QI with (lower panel) or without (upper panel) the perturbation $m_1i\Gamma_1\Gamma_3$ as a function of the intra-cell coupling $\gamma(k_z)$. The vertical direction is added as a visual aid; as such, the narrow loops represent various $3D$ higher-order semimetals. Loop in top panel is a HODSM, loops from \emph{bottom to top} in the bottom panel are $1st$ order, $2nd$ order, and hybrid-order HOWSMs. (c-e) The schematics of three arrangements of Weyl nodes in momentum space corresponding to (c) $H^1$, (d) $H^2$, and (e) $H^3$. Starting from a hybrid-topology WSM in $H^{1,3}$, we can merge a pair of Weyl nodes to realize the $1st$ or $2nd$-order WSMs. The plus and minus signs denote the monopole charges, and the green arrowed lines and cyan dotted lines denote the surface arcs and hinge arcs, respectively.}
    \label{fig:flashpic}
\end{figure}
\begin{figure*}[ht!]
    \centering
    \includegraphics[width=1\textwidth]{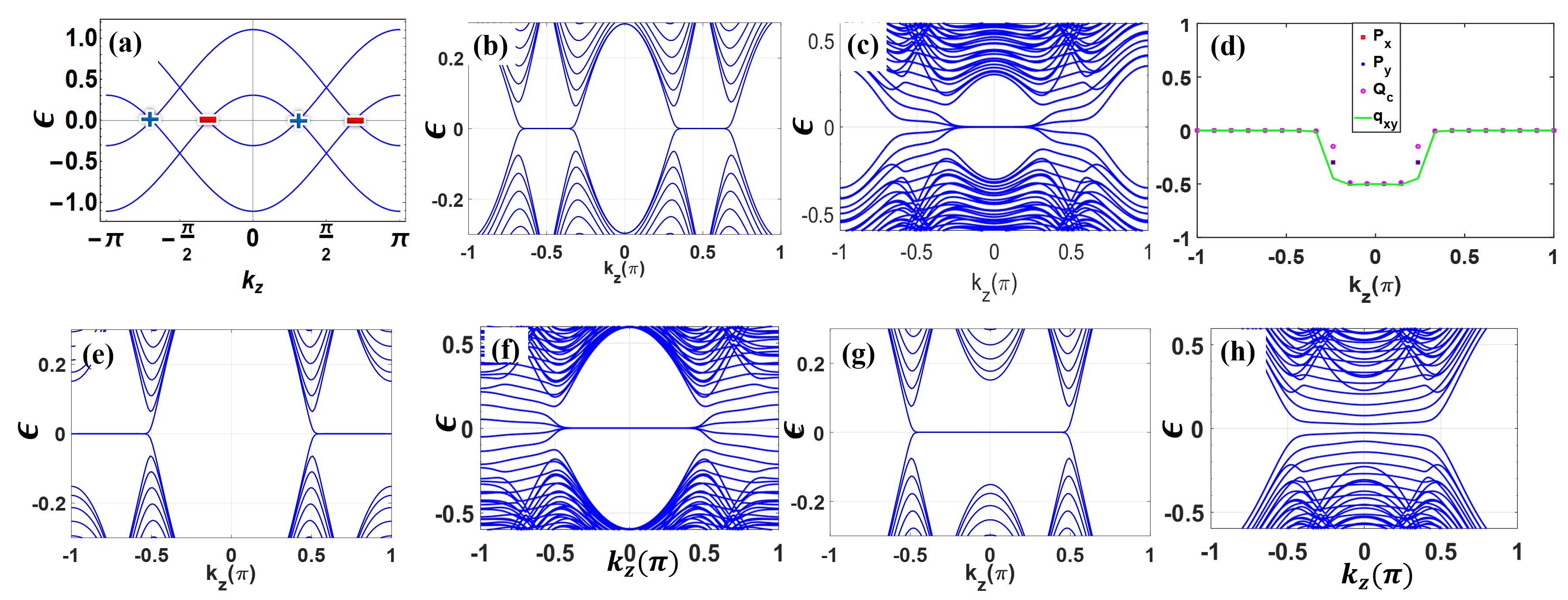}
    \caption{(a) Bulk band structure of $H^1_{HOWSM}$ along $k_z$ for $k_x=k_y=0$. (b) Surface Fermi arcs connecting bulk nodes. (c) hinge Fermi arcs connecting the two middle nodes (higher-order nodes) (d) The combined plot of  $P_x(k_z)$, $P_y(k_z)$, $Q_c(k_z)$ and $q_{xy}(k_z)$ showing the non-zero quantized quadrupole moment corresponding to the region having hinge arcs ($\gamma=-1, m_1=0.4$ used in a-d). (e) Surface and (f) hinge spectrum for a $2nd$-order WSM ($\gamma=-0.7,\,m_1=\frac{0.6}{\sqrt{2}}$, Weyl nodes at $k_z=\pm\frac{\pi}{2}$). (g) Surface and (h) hinge spectrum for a $1st$-order WSM ($\gamma=-1.3,\,m_1=\frac{0.6}{\sqrt{2}}$, Weyl nodes at $k_z=\pm\frac{\pi}{2}$).}
    \label{fig:hyb1st&nd}
\end{figure*}

Here, we will characterize HOWSMs using symmetry considerations, momentum-resolved topological invariants, and their energy spectra.  We present models of HOWSMs which can be easily tuned to $1st$-order, $2nd$-order, and hybrid-order WSMs (the latter where both $1st$ and $2nd$-order nodes coexist). Moreover, our models also allow for tilted Weyl cones that can be tuned to be type-I, type-II or hybrid-tilt \cite{reviewweyl,hybridWeyl}.  In addition, we find that in the presence of magnetic flux a local measurement of charge density can characterize the existence of a HOWSM.
We also show that when a $2nd$-order WSM is gapped out in the bulk by a $1st$-order WSM, the hybridized system is a  hybrid-topology insulating phase that exhibits a coexistence of independent Dirac cones on the surface and flat-band Fermi-arcs on the hinges.  Finally, we propose that circularly polarized light, or an analogous periodic drive, can generate HOWSMs in solid state systems and metamaterials.

\emph{Model and Formalism}.-- We start with a simple model for HODSMs using spinless fermions from Ref. \onlinecite{Lin2017}, whose Bloch Hamiltonian can be written as:
\begin{align}\label{hodsm1}
H_{HODSM}(\vex{k})=&\,\left(\gamma_x +\frac{1}{2}\cos k_z + \cos k_x\right)\Gamma_4+\sin k_x\Gamma_3\cr
+&\,\left(\gamma_y+\frac{1}{2}\cos k_z+\cos k_y\right)\Gamma_2+\sin k_y\Gamma_1,
\end{align}
where $\gamma_{x,y}$ represent the intra-cell coupling along $x, y$,
$\{\Gamma_\alpha\}$ are direct products of Pauli matrices, $\sigma_i,\kappa_i$, following $\Gamma_0=\sigma^3\kappa^0,\Gamma_i=-\sigma^2\kappa^i\,\textrm{for}\,i=1,2,3,$ and $\Gamma_4=\sigma^1\kappa^0$. The corresponding lattice basis configuration is specified in Fig.~\ref{fig:flashpic}(a). Without loss of generality, we have set the amplitudes of inter-cell hoppings to $1,$ and will work in the $C^z_4$ symmetric limit, $\gamma_x=\gamma_y=\gamma$.
In addition to the $C^z_4$ symmetry, the model in Eq.~\eqref{hodsm1} has mirror symmetries $\mathcal{M}_x=\sigma^1\kappa^3$, $\mathcal{M}_y=\sigma^1\kappa^1$, $\mathcal{M}_z=I$, inversion symmetry $\mathcal{I}\equiv\mathcal{M}_x\mathcal{M}_y\mathcal{M}_z=\sigma^0\kappa^2,$ and spinless time-reversal $\mathcal{T}=K$.

Heuristically, Eq.~\eqref{hodsm1} can be viewed as a family of $2D$ quadrupole insulators (QI) having intra-cell coupling amplitudes parameterized by $k_z$: $\gamma(k_z)=\gamma+\frac{1}{2}\cos(k_z)$\cite{Benalcazar2017-1}. The phase diagram of the QI is shown in the upper panel of Fig.~\ref{fig:flashpic}(b), where we see that when $\gamma=-1$ then $-1.5 <\gamma(k_z) <-0.5$, and the $2D$ QI transits between the trivial phase and the topological phase having a quantized quadrupole moment $Q_{xy}=\frac{1}{2}$ as $k_z$ traverses through the Brillouin zone (BZ). This family of $2D$ QIs represents a $3D$ HODSM which hosts two 3D Dirac nodes, one each at $k_z=\pm \arccos(-2(1+\gamma))$, corresponding to the phase transition points (where $\gamma(k_z)=-1$) between a 2D trivial and a 2D quadrupolar phase protected by $C^z_4.$

Interestingly, by breaking certain symmetries, we can split these Dirac nodes and realize various HOWSMs as shown in Fig.~\ref{fig:flashpic}(c-e).
We first consider $H^1=H_{HODSM}+m_1i\Gamma_1\Gamma_3$, which breaks time-reversal symmetry $\mathcal{T}$, $\mathcal{M}_x,$ and $\mathcal{M}_y$, but preserves $C_4^z,$ $\mathcal{M}_x\mathcal{T},$ $\mathcal{M}_y\mathcal{T},$ and $\mathcal{I}$.
The perturbation $i\Gamma_1\Gamma_3$ gaps out/splits the original phase transition point at $\gamma(k_z)= -1$ into two transitions, and generates a new Chern insulator phase $(C=1)$ bounded  by the new critical points at $\gamma(k_z)=-1 \pm\frac{|m_1|}{\sqrt{2}}$\big(see the lower panel of Fig.~\ref{fig:flashpic}(b)\big).
Thus, in the 3D system the bulk Dirac nodes are split into two Weyl nodes.

For this perturbation, the nodes split only along the $k_z$ axis, which is compatible with $C_4^z$. The energies along this axis are $E^{\pm}(k_z)=\pm m_1 \pm \sqrt{\frac{(2+2\gamma+\cos(k_z))^2}{2}}$, indicating four Weyl nodes at $k_z=\pm \arccos[-2\pm \sqrt{2}m_1-2\gamma]$. Owing to the inversion symmetry, the Weyl nodes at opposite $k_z$ carry opposite monopole charge, as denoted in Fig.~\ref{fig:hyb1st&nd}(a). We find that the pair of Weyl nodes at negative $k_z,$ and the pair at positive $k_z$ are both connected through the surface Fermi arcs \big(see Fig.~\ref{fig:hyb1st&nd}(b)\big).
Remarkably, when cutting the surface again to form hinges, the two Weyl nodes closest to $k_z=0$ are connected by additional hinge arcs, as depicted in Fig.~\ref{fig:hyb1st&nd}(c), indicating that they are $2nd$-order Weyl nodes that act as the critical point (as a function of $k_z$) between a Chern insulator, and a QI having vanishing Chern number.
Since this HOWSM consists of both $1st$ and $2nd$-order Weyl nodes (as of Fig.~\ref{fig:hyb1st&nd}(a-d)), we can call it a \textit{hybrid-order Weyl semimetal}.

To characterize the higher-order topology we can calculate the bulk quadrupole moment $q_{xy}(k_z)$ for each $k_z$ slice using $q_{xy}(k_z)\equiv (P_x(k_z) + P_y(k_z) - Q_c(k_z))\mod 1$~\cite{Lin2017,Benalcazar2017-1, Benalcazar2017-2}, where $P_x(k_z)\big(P_y(k_z)\big)$ is the $x(y)$-component of the polarization localized at the $y(x)$-surface, and $Q_c(k_z)$ is the corner charge.
Fig.~\ref{fig:hyb1st&nd}(d) shows that $q_{xy}(k_z)$ is quantized to a non-trivial value of $-\frac{1}{2}$ for the $k_z$ range where the hinge arcs exist.
Therefore, the hinge states of $H^{1}$ can be captured by a second-order topological invariant.
We can merge the $1st$-order ($2nd$-order) Weyl nodes independently in $H^1$ by increasing (decreasing) $\gamma$, to transition the system to a $2nd$-order WSM ($1st$-order WSM)\big(see Fig.~\ref{fig:hyb1st&nd}(e-h)\big).
The merging of Weyl nodes can be understood using the phase diagram shown in the lower panel of Fig.~\ref{fig:flashpic}(b).
By tuning $\gamma$, the trajectory of $\gamma(k_z)$ shifts horizontally; when the trajectory crosses only critical points separating a trivial phase and a $C=1$ phase, the corresponding $3D$ WSM hosts $1st$-order Weyl nodes, while for critical points separating a $C=1$ and the QI phase, the $3D$ WSM hosts $2nd$-order Weyl nodes.

Unlike the $1st$-order Weyl nodes, which separate two phases having different (possibly both non-zero) Chern numbers, the $2nd$-order Weyl node requires one of the two separated phases to have a non-zero quantized higher-order invariant, which usually necessitates the Chern number to vanish in that phase.
In $C_4$ symmetric systems, both the Chern number (modulo $4$) and the quadrupole moment can be determined by the symmetry representations of the $C_4^z$ operator formed by occupied bands at the high symmetry points in the BZ\cite{fang2012bulk,Classification2014Wladimir,Benalcazar2017-2}. These symmetry indicators can therefore be used as a bulk diagnosis for $2nd$-order Weyl nodes (see Supplement \cite{sm}). However, it is important to note that in order to characterize the $2nd$-order nodes, the symmetry indicators of just the two  crossing bands \emph{are not sufficient}. Another way to understand this is to notice that the low-energy  descriptions (e.g., $k\cdot P$) of $1st$- and $2nd$-order nodes are identical, therefore the other occupied bands need to be considered to determine the order of the node.

\indent
From Fig.~\ref{fig:flashpic}(b), we see that hinge arcs in the HOWSM emanate from the projections of the bulk nodes on the hinges. We expect this since $k_z$-slices having $\frac{1}{2}$ quadrupole moment host corner states when truncating in $\hat{x}$ and $\hat{y}$. However, unlike the surface arcs, the hinge arcs are protected by $C_4^z$ symmetry and, while their degeneracy is protected, their energies can be pushed outside of the midgap region.  It may be difficult to  precisely control the energies of the hinge modes in electronic materials, perhaps making their observation more challenging. However, it is typically straightforward to manipulate the energies of boundary modes in metamaterials contexts by modifying the effective boundary conditions. Additionally, the attachment of the hinge arc to a more stable bulk feature, i.e., the Weyl node, may also aid in their observation.
Finally, even if the hinge modes are completely removed one can still generically use fractional charge density on the hinges as a measure of the higher order topology\cite{benalcazar2019,Peterson1114,peterson2020observation}.

\indent As a second example of a HOWSM  we can consider a model in which we add the same perturbation in $H^1,$ but with an explicit momentum-dependence: $H^{2}(\vex{k})=H_{HODSM}(\vex{k})+m_2\sin(k_z)i\Gamma_1\Gamma_3$. This has the effect of restoring $\mathcal{T}$ while maintaining $C^z_4$.  In contrast to $\mathcal{T}$-broken WSMs, in the presence of $\mathcal{T}$ the minimum number of nodes is four, and so the $1st$ or $2nd$-order pairs cannot be merged separately. Instead, by tuning $\gamma,$ all of the nodes merge to gap out, as illustrated in Fig.~\ref{fig:flashpic}(d). Similar to $H^1$, the symmetry representations of $C^z_4$ can be used for a bulk diagnosis of the $2nd$-order nodes, however, knowing the states at a single band crossing is unfortunately not enough to uniquely determine the order of the node.

\indent So far, we have realized two HOWSMs with $C_{4}^z$ symmetry where all Weyl nodes are aligned along the high symmetry line $(k_x,k_y)=(0,0)$.
We now show that one can realize HOWSMs where nodes are split in other planes by applying a different perturbation to Eq.~\ref{hodsm1}.
Consider $H^{3}(\vex{k})=H_{HODSM}(\vex{k})+m_3\sin(k_z)i\Gamma_2\Gamma_3$, which explicitly breaks $\mathcal{M}_z$ and $\mathcal{T},$ but preserves $\mathcal{I}$.
Since $\mathcal{M}_x$ is also broken, but $\mathcal{M}_y$ is preserved, the Dirac nodes split in only the $k_y$-$k_z$ plane. As a result, there are now surface Fermi arcs connecting nodes split in the $k_y$ direction. Since we have $\mathcal{I}$ symmetry the four-nodes form two parallel dipoles of monopole charges \big(see  Fig.~\ref{fig:flashpic}(e)\big). Unlike $H^{1,2}$,   the surface and hinge arcs of $H^{3}$ are perpendicular to each other (see Fig.~\ref{fig:flashpic}(d, e) for a schematic illustration and \cite{sm} for a plot of the spectrum). We find that the higher-order topology can still be characterized using the quadrupole moment, which is well-defined since $M_y$ quantizes the quadrupole moment, while the polarization is quantized by $\mathcal{I},$ and vanishes for this parameter regime. 
Finally, in addition to these three examples, we note that we considered models of HOWSMs  having various arrangements of Weyl nodes possessing hinge states which belong to the category of ``extrinsic HOWSMs" \cite{extrinsichoti1}. Some examples of these models discussed in SM \cite{sm}.

\emph{Hybridizing $1st$- and $2nd$- order WSMs.}-- We can use these HOWSM models to generate an interesting type of hybrid-order topological insulator.
Since the $2nd$-order nodes can manifest boundary features beyond surface arcs, it is natural to ask what will happen when we couple Weyl nodes of different orders.
As a proof of concept, let us consider coupling the $2nd$-order WSM in Fig.~\ref{fig:hyb1st&nd}(e, f) with the $1st$-order WSM in Fig.~\ref{fig:hyb1st&nd}(g, h). Both phases are generated from the same model, but have different values of $\gamma.$
In order to gap out the bulk nodes, we change $\gamma(k_z)$ in Eq.~\eqref{hodsm1} to $\gamma(k_z+\pi)$ for the $1st$-order WSM, and reverse the sign of the perturbation $m_1$ such that its surface Fermi arcs overlap with the surface arcs in the $2nd$-order WSM in Fig.~\ref{fig:hyb1st&nd}(e). Then, the Hamiltonian of the hybridized system is,
\begin{align}
   h_{1\&2}=\left[
  \begin{array}{cc}
  H_{WSM}(-m_1, k_z+\pi) & t\hat{T}\\
  t^\ast \hat{T}^{\dagger} & H^1_{HOWSM}(m_1, k_z)
  \end{array}
\right],
\end{align}
where $\hat{T}$ is the hybridization matrix which couples the two Weyl phases, and $t$ is the coupling strength.

\begin{figure}[t]
    \includegraphics[width=0.45\textwidth]{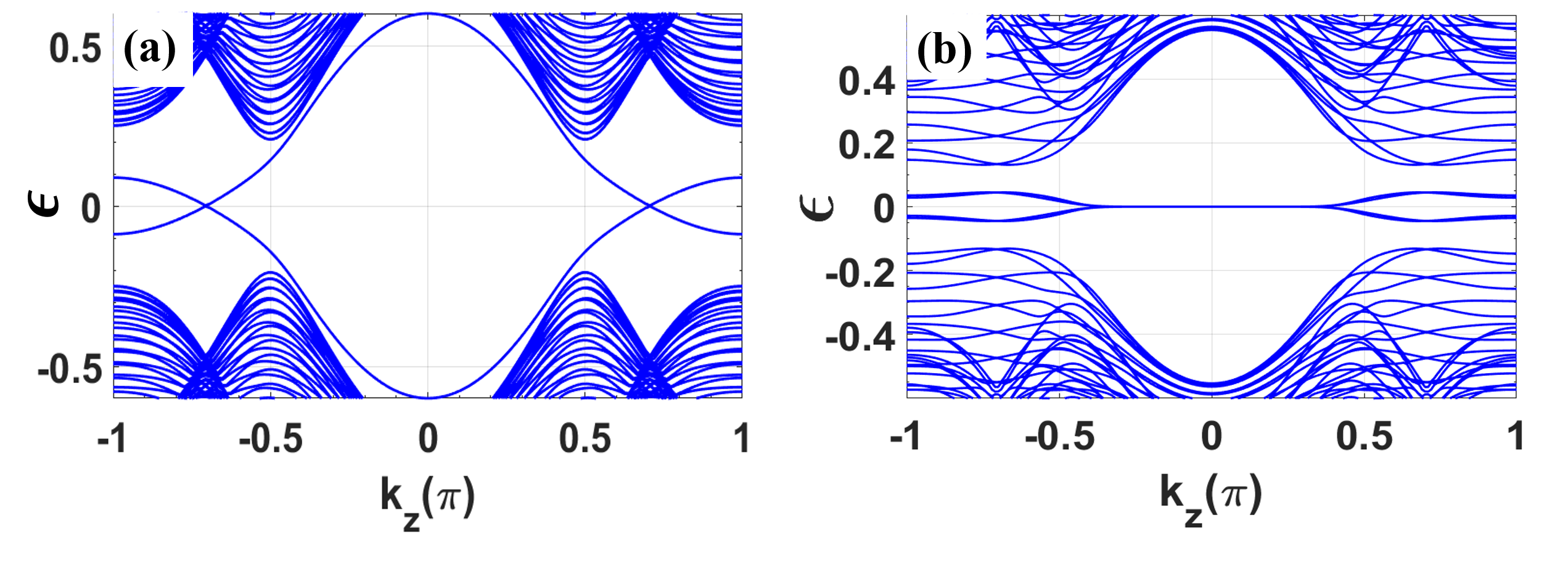}
     \includegraphics[width=0.5\textwidth]{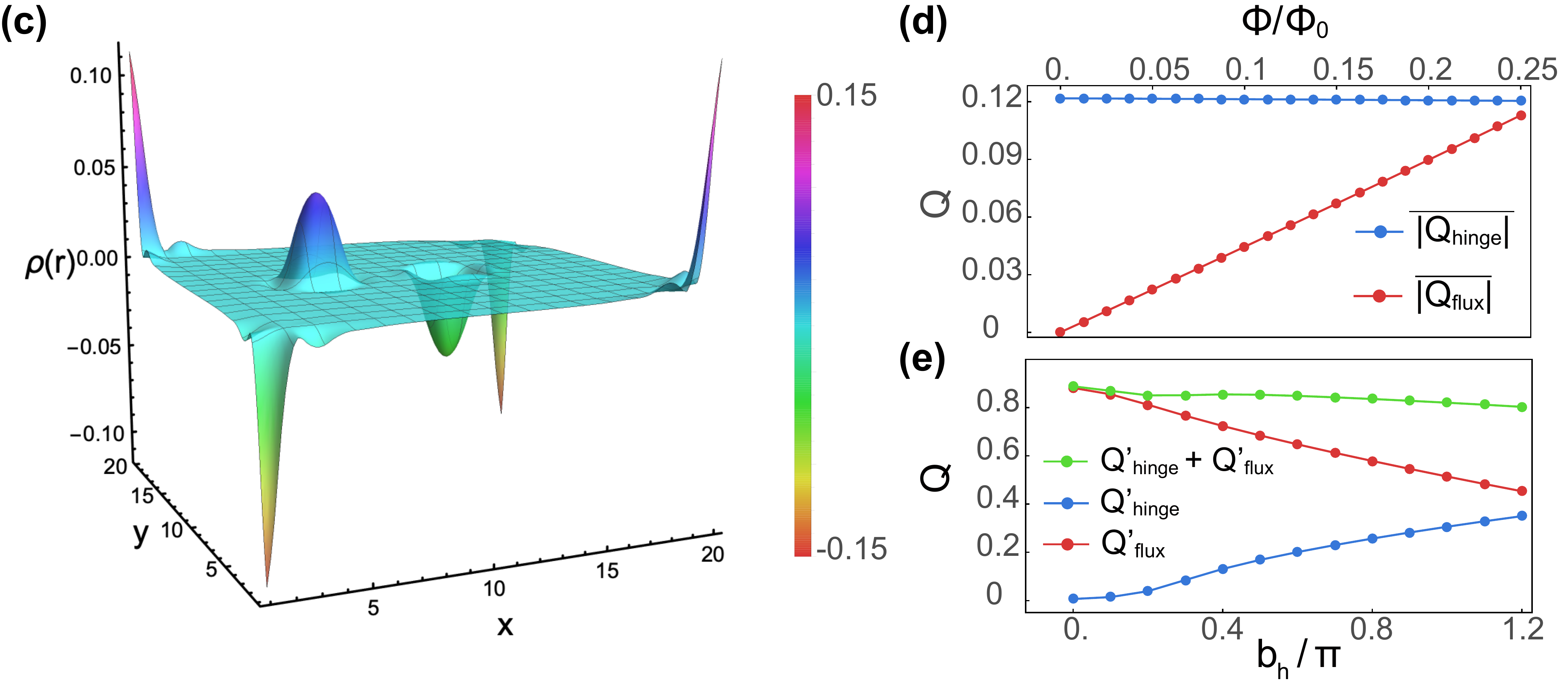}
    \caption{(a) The surface and (b) hinge spectrum of the hybrid topological insulator $h_{1\&2}$. Electromagnetic response of HOWSMs: (c) Charge density at half-filling for $H_{HOWSM}^1$ having localized magnetic flux along $z$ and open boundary condition along $x$ and $y$. (d, e) Change in hinge charge and flux charge by (d) tuning magnetic field and (e) the length of the hinge arc. $Q_{hinge}'=\frac{2Q_{hinge}}{e}, Q_{flux}'=Q_{flux}\frac{\Phi_0}{e\Phi_z}$.}
    \label{fig:hyb&flux}
\end{figure}

To form an insulating phase we need to choose $\hat{T}$ such that the bulk nodes are hybridized and gapped. We find that the only $\hat{T}$ that can fully gap out the bulk nodes is $\hat{T}=\sigma^0\kappa^2=i\Gamma_1\Gamma_3$.
As shown in Fig.~\ref{fig:hyb&flux}(a, b), this insulating phase of $h_{1\&2}$ is a hybrid-order topological insulator with coexisting surface Dirac cone pairs and flat-band hinge states where the former are protected by the combination of $\mathcal{M}_x\mathcal{T}$ and $\mathcal{M}_y\mathcal{T}$ symmetries, and the latter protected by $C_4^z$ as before.
Remarkably, the hinge states do not originate from the surface cones as occurs in graphene, for example, or in the quadrupolar surface semimetals shown in \cite{Lin2017}. Instead, they are protected by independent symmetries and can be gapped out independently. For example, one can break $C_4^z$ while preserving $\mathcal{M}_x\mathcal{T},\mathcal{M}_y\mathcal{T}$ to remove the hinge states (see \cite{sm} for details). This  leads to a type of  hybrid-order topology which to our knowledge has not been reported elsewhere. We can also use this result to identify HOWSMs. If one couples two Weyl systems with just $1st$-order (or just $2nd$-order) Weyl nodes and gap out the bulk bands, the resulting insulator is either a stack of Chern insulators having $C=1$ for each $k_z$-slice, or a trivial insulator. In the case of coupling two $2nd$-order WSMs, there is another possibility where there could instead exist flat band hinge states throughout the whole BZ (or associated fractional charge density on the hinges). The unusual hybrid-order insulator arising form the combination of $1st$-order and $2nd$-order WSMs allows us to use the coupling to conventional WSMs as a mean to detect the higher-order topology of a $2nd$-order WSM. It would also be particularly interesting to consider a $2nd$-order WSM gapped by a charge density wave analogous to recent work on axionic insulators \cite{cdw1,cdw2,cdw3,cdw4,cdw5,wieder2020dynamical}.
As mentioned above our results predict such a system can exhibit  hinge states throughout the whole BZ, we will leave a complete analysis for the future.

\emph{Electromagnetic response of $2nd$-order Weyl nodes.}-- We will now show that $2nd$-order WSMs can show an interesting electromagnetic response as a result of the coexistence of surface and hinge Fermi arcs.
Let us consider the case of $H^1_{HOWSM}$ in a regime where we have two $2nd$-order Weyl nodes located on the $k_z$-axis.
It is known that the charge response in a conventional WSMs manifests as $\rho=\frac{e^2\vex{b}_s\cdot\vex{B}}{4\pi^2}$\cite{vazifeh,burkovPRL2011,PhysRevB.83.205101}, where $\vex{b}_s$ is the vector connecting the two Weyl nodes.
%
%
Hence, the charge per layer along $\hat{z}$ in the presence of a magnetic flux is $|Q_f| = \vert\frac{ b_s}{2\pi}\frac{e \Phi_z}{\Phi_0}\vert$ which is proportional to the flux strength ($\Phi_z/\Phi_0=e\Phi_z/h$) for the applied magnetic field in the $z$-direction, and the nodal separation along $k_z$ \cite{vazifeh}.
Simultaneously, for a system with open boundary conditions in the $x$ and $y$ directions we also expect localized fractional charge at the hinges parallel to $\hat{z}$, whose value is proportional to the (momentum-space) length of the hinge arcs, $b_h$: $|Q_h| =\vert\frac{b_h}{2\pi}  \frac{e}{2}\vert$. In the case of minimal models having only  two ($2nd$-order) Weyl nodes separated along the $k_z$-axis, the length of hinge and the surface arcs together must span the whole BZ, i.e., $b_s + b_h=2\pi$ or equivalently, $|Q_f|\frac{\Phi_0}{e\Phi_z} + |Q_h|\frac{2}{e} = 1$. For example, if we insert one flux quantum this yields the remarkable result $|Q_f|+2|Q_h|=e$ independent of the other details of the system.

To confirm these charge distributions numerically let us take the model of $H_{HOWSM}^1$ in Fig.~\ref{fig:hyb1st&nd}(e,f) and insert two oppositely oriented flux lines localized in the $xy$-plane. The charge density at half-filling in Fig.~\ref{fig:hyb&flux}(a) shows the charge accumulation at both the fluxes and the hinges.
While the charge bound to the flux is proportional to the external flux, the hinge charge is insensitive to it, as shown in Fig.~\ref{fig:hyb&flux}(d).
However, due to the constraint on $b_s$ and $b_h$ for this two-node WSM there is a competition between the hinge and flux charges as shown in Fig.~\ref{fig:hyb&flux}(e). Namely, for a fixed amount of flux, increasing the length of the hinge arcs in $H_{HOWSM}^1$, increases $Q_{h}$ but decreases $Q_f$; while the weighted sum of $Q_{h}$ and $Q_f$ remains constant during this process. Thanks to the recent developments in charge measurement analogs in metamaterials \cite{Peterson1114} we expect this  property can serve as an experimental indicator for the HOWSM phase in those contexts, and possibly in electronic materials.

\begin{figure}[t]
    \includegraphics[width=0.495\textwidth]{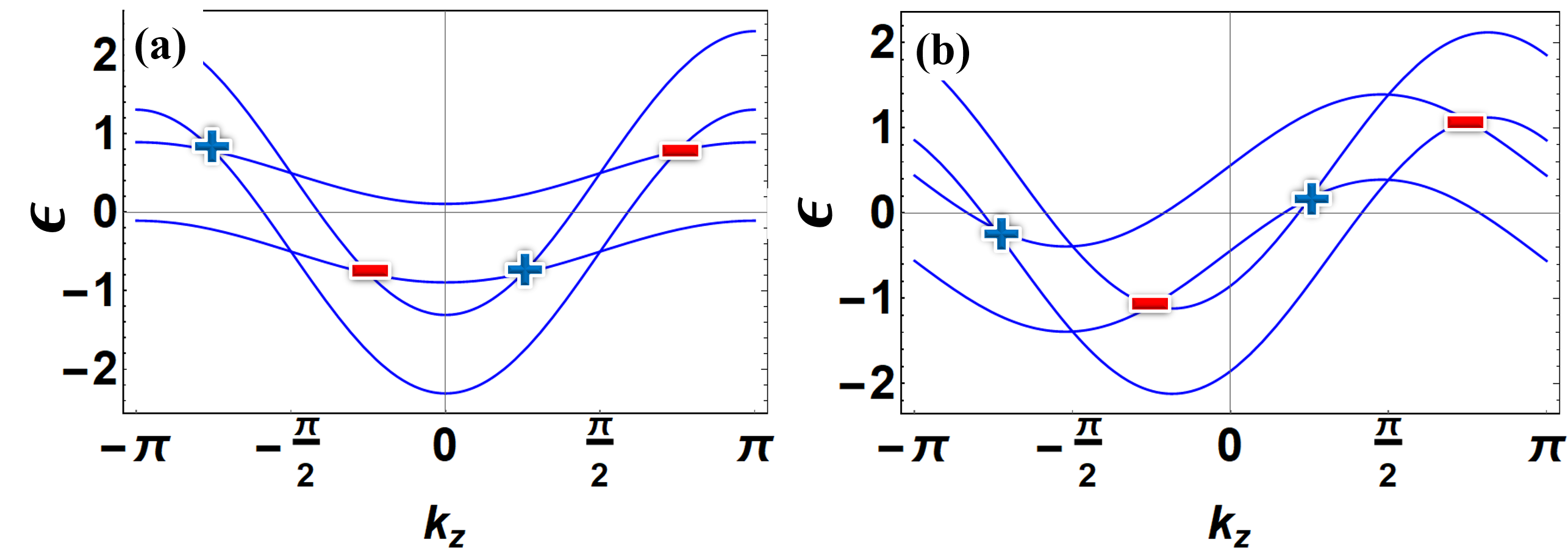}
    \caption{Bulk band structure showing (a) type-II hybrid-order HOWSM at $\theta=\pi/2$ (b) and hybrid-tilt, hybrid-order HOWSM for $\theta=\pi/5$ ($\gamma=-1,\,m=0.5,\alpha=1.1$).}
    \label{fig:typeII-hybrid}
\end{figure}

\indent\emph{Type-II and Hybrid higher-order Weyl semimetals}.--
Conventional type-II WSMs have dispersion that is strongly anisotropic around the Weyl nodes such that its slope changes sign along some directions~\cite{reviewweyl,Soluyanov2015}.
Additionally, when a pair of Weyl nodes consists of a type-I (no tilt) and a type-II Weyl node, they form a so-called hybrid-tilt Weyl phase \cite{hybridWeyl,GhorashiFloquet2}.
We can extend the models introduced here (we only show this for the model $H^1_{HOWSM}$) to introduce higher-order type-II and hybrid-tilt Weyl phases. This can be done by including a term proportional to the identity matrix: $\alpha\sin(k_z-\theta)\mathbb{I}_4$. For $\theta=\pi/2$, and for a sufficiently strong $\alpha,$ the Weyl nodes undergo a transition to type-II nodes (Fig.~\ref{fig:typeII-hybrid}(a)). Furthermore, by tuning $\theta$ to a small value, the nodal tilt can be tuned to form a hybrid-tilt Weyl phase. Fig.~\ref{fig:typeII-hybrid}(b), shows the bulk bandstructure of the $H^1_{HOWSM}$, at e.g., $\theta=\pi/5$ where all the Weyl pairs (both $1st$ and $2nd$-order) form hybrid-tilt Weyl phases. On the other hand, as discussed previously, by adjusting parameters such as $\gamma$ and $m_1$ we can tune the system to $1st$-order or $2nd$-order WSMs. Therefore, this model can realize a complete set of type-I, type-II and hybrid-tilt phases having $1st$-, $2nd$- and hybrid-order topology.

\indent \emph{Possible experimental realization}.--The 2D QI model proposed in Ref.\cite{Benalcazar2017-1} is already experimentally realized in multiple  metamaterial contexts\cite{Peterson2018,Noh2018,Imhof2018}.
In light of these developments, building a metamaterial HODSM as in Eq.~\eqref{hodsm1}  constructed from a stack of 2D QI layers is feasible. Moreover, solid-state candidates were recently proposed to be HODSMs, e.g., the room- ($\alpha$) and intermediate-temperature ($\alpha"$) phases of $\text{Cd}_3\text{As}_2$, KMgBi, and  rutile-structure ($\beta'$-) $\text{PtO}_2$ \cite{Wieder2020,PhysRevLett.124.156601}.
Now let us show that the HOWSMs introduced in this work can be dynamically generated from a HODSM through periodic driving protocols. Periodic driving has already proven  useful in inducing Weyl phases in conventional topological semimetals \cite{hubener2017creating,GhorashiFloquet1,GhorashiFloquet2,oka2019floquet}.
For circularly polarized light (or an analogous driving field) of the form $\vex{A}(t)=(A_0\eta \cos(\omega t), A_0\sin(\omega t), 0)$, in the high-frequency limit we obtain a time-independent Hamiltonian:
$h^{W1}_{eff}(\vex{k})\,=H_{HODSM}(\vex{k})+\frac{iA^2_0\eta}{2\omega}\bigg(\sin(k_x)\sin(ky)[\Gamma_2,\Gamma_4]-[\Gamma_1,\Gamma_4]\sin(k_x)-[\Gamma_2,\Gamma_3]\sin(k_y)+[\Gamma_1, \Gamma_3]\bigg),$
where $A_0\propto E/\omega,$ $E$ is the electric field, and $\eta=\pm 1$ is the chirality of the light.
At $k_x=k_y=0$, the effect of light is $\propto i\Gamma_1\Gamma_3$, which can split the two Dirac nodes into four Weyl nodes on the $k_z$ axis (see \cite{sm} for the details). Therefore, we expect that driving the HODSM in Eq.\eqref{hodsm1} using circular polarized light can produce the same physics as $H_{HOWSM}^1$. \\
\indent  Therefore, we expect that the physics discussed in this work can help lead to the discovery as well as the understanding of HOWSMs.

\emph{Acknowledgement}.---S.A.A.G. acknowledges support from ARO (Grant No. W911NF-18-1-0290) and NSF (Grant No. DMR1455233). TL
thanks the US National Science Foundation (NSF) MRSEC
program under NSF Award Number DMR-1720633 (SuperSEED) for support. TLH thanks the US Office of Naval Research (ONR) Multidisciplinary University Research Initiative (MURI) grant N00014-20-1-2325 on Robust Photonic Materials with High-Order Topological Protection.


\begin{thebibliography}{67}%
\makeatletter
\providecommand \@ifxundefined [1]{%
 \@ifx{#1\undefined}
}%
\providecommand \@ifnum [1]{%
 \ifnum #1\expandafter \@firstoftwo
 \else \expandafter \@secondoftwo
 \fi
}%
\providecommand \@ifx [1]{%
 \ifx #1\expandafter \@firstoftwo
 \else \expandafter \@secondoftwo
 \fi
}%
\providecommand \natexlab [1]{#1}%
\providecommand \enquote  [1]{``#1''}%
\providecommand \bibnamefont  [1]{#1}%
\providecommand \bibfnamefont [1]{#1}%
\providecommand \citenamefont [1]{#1}%
\providecommand \href@noop [0]{\@secondoftwo}%
\providecommand \href [0]{\begingroup \@sanitize@url \@href}%
\providecommand \@href[1]{\@@startlink{#1}\@@href}%
\providecommand \@@href[1]{\endgroup#1\@@endlink}%
\providecommand \@sanitize@url [0]{\catcode `\\12\catcode `\$12\catcode
  `\&12\catcode `\#12\catcode `\^12\catcode `\_12\catcode `\%12\relax}%
\providecommand \@@startlink[1]{}%
\providecommand \@@endlink[0]{}%
\providecommand \url  [0]{\begingroup\@sanitize@url \@url }%
\providecommand \@url [1]{\endgroup\@href {#1}{\urlprefix }}%
\providecommand \urlprefix  [0]{URL }%
\providecommand \Eprint [0]{\href }%
\providecommand \doibase [0]{http://dx.doi.org/}%
\providecommand \selectlanguage [0]{\@gobble}%
\providecommand \bibinfo  [0]{\@secondoftwo}%
\providecommand \bibfield  [0]{\@secondoftwo}%
\providecommand \translation [1]{[#1]}%
\providecommand \BibitemOpen [0]{}%
\providecommand \bibitemStop [0]{}%
\providecommand \bibitemNoStop [0]{.\EOS\space}%
\providecommand \EOS [0]{\spacefactor3000\relax}%
\providecommand \BibitemShut  [1]{\csname bibitem#1\endcsname}%
\let\auto@bib@innerbib\@empty
\bibitem [{\citenamefont {Benalcazar}\ \emph
  {et~al.}(2017{\natexlab{a}})\citenamefont {Benalcazar}, \citenamefont
  {Bernevig},\ and\ \citenamefont {Hughes}}]{Benalcazar2017-1}%
  \BibitemOpen
  \bibfield  {author} {\bibinfo {author} {\bibfnamefont {Wladimir~A}\
  \bibnamefont {Benalcazar}}, \bibinfo {author} {\bibfnamefont {B~Andrei}\
  \bibnamefont {Bernevig}}, \ and\ \bibinfo {author} {\bibfnamefont {Taylor~L}\
  \bibnamefont {Hughes}},\ }\bibfield  {title} {\enquote {\bibinfo {title}
  {{Quantized electric multipole insulators.}}}\ }\href {\doibase
  10.1126/science.aah6442} {\bibfield  {journal} {\bibinfo  {journal}
  {Science}\ }\textbf {\bibinfo {volume} {357}},\ \bibinfo {pages} {61--66}
  (\bibinfo {year} {2017}{\natexlab{a}})}\BibitemShut {NoStop}%
\bibitem [{\citenamefont {Benalcazar}\ \emph
  {et~al.}(2017{\natexlab{b}})\citenamefont {Benalcazar}, \citenamefont
  {Bernevig},\ and\ \citenamefont {Hughes}}]{Benalcazar2017-2}%
  \BibitemOpen
  \bibfield  {author} {\bibinfo {author} {\bibfnamefont {Wladimir~A}\
  \bibnamefont {Benalcazar}}, \bibinfo {author} {\bibfnamefont {B~Andrei}\
  \bibnamefont {Bernevig}}, \ and\ \bibinfo {author} {\bibfnamefont {Taylor~L}\
  \bibnamefont {Hughes}},\ }\bibfield  {title} {\enquote {\bibinfo {title}
  {{Selected for a Viewpoint in Physics Electric multipole moments, topological
  multipole moment pumping, and chiral hinge states in crystalline
  insulators}},}\ }\href {\doibase 10.1103/PhysRevB.96.245115} {\bibfield
  {journal} {\bibinfo  {journal} {Physical Review B}\ }\textbf {\bibinfo
  {volume} {96}},\ \bibinfo {pages} {245115} (\bibinfo {year}
  {2017}{\natexlab{b}})}\BibitemShut {NoStop}%
\bibitem [{\citenamefont {Song}\ \emph {et~al.}(2017)\citenamefont {Song},
  \citenamefont {Fang},\ and\ \citenamefont {Fang}}]{Song2017}%
  \BibitemOpen
  \bibfield  {author} {\bibinfo {author} {\bibfnamefont {Zhida}\ \bibnamefont
  {Song}}, \bibinfo {author} {\bibfnamefont {Zhong}\ \bibnamefont {Fang}}, \
  and\ \bibinfo {author} {\bibfnamefont {Chen}\ \bibnamefont {Fang}},\
  }\bibfield  {title} {\enquote {\bibinfo {title}
  {$(d\ensuremath{-}2)$-dimensional edge states of rotation symmetry protected
  topological states},}\ }\href {\doibase 10.1103/PhysRevLett.119.246402}
  {\bibfield  {journal} {\bibinfo  {journal} {Phys. Rev. Lett.}\ }\textbf
  {\bibinfo {volume} {119}},\ \bibinfo {pages} {246402} (\bibinfo {year}
  {2017})}\BibitemShut {NoStop}%
\bibitem [{\citenamefont {Schindler}\ \emph
  {et~al.}(2018{\natexlab{a}})\citenamefont {Schindler}, \citenamefont {Cook},
  \citenamefont {Vergniory}, \citenamefont {Wang}, \citenamefont {Parkin},
  \citenamefont {Bernevig},\ and\ \citenamefont {Neupert}}]{Schindler2018-1}%
  \BibitemOpen
  \bibfield  {author} {\bibinfo {author} {\bibfnamefont {Frank}\ \bibnamefont
  {Schindler}}, \bibinfo {author} {\bibfnamefont {Ashley~M.}\ \bibnamefont
  {Cook}}, \bibinfo {author} {\bibfnamefont {Maia~G.}\ \bibnamefont
  {Vergniory}}, \bibinfo {author} {\bibfnamefont {Zhijun}\ \bibnamefont
  {Wang}}, \bibinfo {author} {\bibfnamefont {Stuart S.~P.}\ \bibnamefont
  {Parkin}}, \bibinfo {author} {\bibfnamefont {B.~Andrei}\ \bibnamefont
  {Bernevig}}, \ and\ \bibinfo {author} {\bibfnamefont {Titus}\ \bibnamefont
  {Neupert}},\ }\bibfield  {title} {\enquote {\bibinfo {title} {{Higher-order
  topological insulators}},}\ }\href {\doibase 10.1126/sciadv.aat0346}
  {\bibfield  {journal} {\bibinfo  {journal} {Science Advances}\ }\textbf
  {\bibinfo {volume} {4}},\ \bibinfo {pages} {eaat0346} (\bibinfo {year}
  {2018}{\natexlab{a}})}\BibitemShut {NoStop}%
\bibitem [{\citenamefont {Langbehn}\ \emph {et~al.}(2017)\citenamefont
  {Langbehn}, \citenamefont {Peng}, \citenamefont {Trifunovic}, \citenamefont
  {von Oppen},\ and\ \citenamefont {Brouwer}}]{Langbehn2017}%
  \BibitemOpen
  \bibfield  {author} {\bibinfo {author} {\bibfnamefont {Josias}\ \bibnamefont
  {Langbehn}}, \bibinfo {author} {\bibfnamefont {Yang}\ \bibnamefont {Peng}},
  \bibinfo {author} {\bibfnamefont {Luka}\ \bibnamefont {Trifunovic}}, \bibinfo
  {author} {\bibfnamefont {Felix}\ \bibnamefont {von Oppen}}, \ and\ \bibinfo
  {author} {\bibfnamefont {Piet~W.}\ \bibnamefont {Brouwer}},\ }\bibfield
  {title} {\enquote {\bibinfo {title} {Reflection-symmetric second-order
  topological insulators and superconductors},}\ }\href {\doibase
  10.1103/PhysRevLett.119.246401} {\bibfield  {journal} {\bibinfo  {journal}
  {Phys. Rev. Lett.}\ }\textbf {\bibinfo {volume} {119}},\ \bibinfo {pages}
  {246401} (\bibinfo {year} {2017})}\BibitemShut {NoStop}%
\bibitem [{\citenamefont {Benalcazar}\ \emph {et~al.}(2019)\citenamefont
  {Benalcazar}, \citenamefont {Li},\ and\ \citenamefont
  {Hughes}}]{benalcazar2019}%
  \BibitemOpen
  \bibfield  {author} {\bibinfo {author} {\bibfnamefont {Wladimir~A.}\
  \bibnamefont {Benalcazar}}, \bibinfo {author} {\bibfnamefont {Tianhe}\
  \bibnamefont {Li}}, \ and\ \bibinfo {author} {\bibfnamefont {Taylor~L.}\
  \bibnamefont {Hughes}},\ }\bibfield  {title} {\enquote {\bibinfo {title}
  {Quantization of fractional corner charge in ${C}_{n}$-symmetric higher-order
  topological crystalline insulators},}\ }\href {\doibase
  10.1103/PhysRevB.99.245151} {\bibfield  {journal} {\bibinfo  {journal} {Phys.
  Rev. B}\ }\textbf {\bibinfo {volume} {99}},\ \bibinfo {pages} {245151}
  (\bibinfo {year} {2019})}\BibitemShut {NoStop}%
\bibitem [{\citenamefont {Ghorashi}\ \emph
  {et~al.}(2019{\natexlab{a}})\citenamefont {Ghorashi}, \citenamefont
  {Hughes},\ and\ \citenamefont {Rossi}}]{ghorashi2019vortex}%
  \BibitemOpen
  \bibfield  {author} {\bibinfo {author} {\bibfnamefont {Sayed Ali~Akbar}\
  \bibnamefont {Ghorashi}}, \bibinfo {author} {\bibfnamefont {Taylor~L.}\
  \bibnamefont {Hughes}}, \ and\ \bibinfo {author} {\bibfnamefont {Enrico}\
  \bibnamefont {Rossi}},\ }\href@noop {} {\enquote {\bibinfo {title} {Vortex
  and surface phase transitions in superconducting higher-order topological
  insulators},}\ } (\bibinfo {year} {2019}{\natexlab{a}}),\ \Eprint
  {http://arxiv.org/abs/1909.10536} {arXiv:1909.10536 [cond-mat.supr-con]}
  \BibitemShut {NoStop}%
\bibitem [{\citenamefont {Li}\ \emph {et~al.}(2020{\natexlab{a}})\citenamefont
  {Li}, \citenamefont {Zhu}, \citenamefont {Benalcazar},\ and\ \citenamefont
  {Hughes}}]{tianhe1}%
  \BibitemOpen
  \bibfield  {author} {\bibinfo {author} {\bibfnamefont {Tianhe}\ \bibnamefont
  {Li}}, \bibinfo {author} {\bibfnamefont {Penghao}\ \bibnamefont {Zhu}},
  \bibinfo {author} {\bibfnamefont {Wladimir~A.}\ \bibnamefont {Benalcazar}}, \
  and\ \bibinfo {author} {\bibfnamefont {Taylor~L.}\ \bibnamefont {Hughes}},\
  }\bibfield  {title} {\enquote {\bibinfo {title} {Fractional disclination
  charge in two-dimensional ${C}_{n}$-symmetric topological crystalline
  insulators},}\ }\href {\doibase 10.1103/PhysRevB.101.115115} {\bibfield
  {journal} {\bibinfo  {journal} {Phys. Rev. B}\ }\textbf {\bibinfo {volume}
  {101}},\ \bibinfo {pages} {115115} (\bibinfo {year}
  {2020}{\natexlab{a}})}\BibitemShut {NoStop}%
\bibitem [{\citenamefont {Serra-Garcia}\ \emph {et~al.}(2018)\citenamefont
  {Serra-Garcia}, \citenamefont {Peri}, \citenamefont {S{\"{u}}sstrunk},
  \citenamefont {Bilal}, \citenamefont {Larsen}, \citenamefont {Villanueva},\
  and\ \citenamefont {Huber}}]{Serra-Garcia2018}%
  \BibitemOpen
  \bibfield  {author} {\bibinfo {author} {\bibfnamefont {Marc}\ \bibnamefont
  {Serra-Garcia}}, \bibinfo {author} {\bibfnamefont {Valerio}\ \bibnamefont
  {Peri}}, \bibinfo {author} {\bibfnamefont {Roman}\ \bibnamefont
  {S{\"{u}}sstrunk}}, \bibinfo {author} {\bibfnamefont {Osama~R.}\ \bibnamefont
  {Bilal}}, \bibinfo {author} {\bibfnamefont {Tom}\ \bibnamefont {Larsen}},
  \bibinfo {author} {\bibfnamefont {Luis~Guillermo}\ \bibnamefont
  {Villanueva}}, \ and\ \bibinfo {author} {\bibfnamefont {Sebastian~D.}\
  \bibnamefont {Huber}},\ }\bibfield  {title} {\enquote {\bibinfo {title}
  {{Observation of a phononic quadrupole topological insulator}},}\ }\href
  {\doibase 10.1038/nature25156} {\bibfield  {journal} {\bibinfo  {journal}
  {Nature}\ }\textbf {\bibinfo {volume} {555}},\ \bibinfo {pages} {342--345}
  (\bibinfo {year} {2018})}\BibitemShut {NoStop}%
\bibitem [{\citenamefont {Peterson}\ \emph {et~al.}(2018)\citenamefont
  {Peterson}, \citenamefont {Benalcazar}, \citenamefont {Hughes},\ and\
  \citenamefont {Bahl}}]{Peterson2018}%
  \BibitemOpen
  \bibfield  {author} {\bibinfo {author} {\bibfnamefont {Christopher~W.}\
  \bibnamefont {Peterson}}, \bibinfo {author} {\bibfnamefont {Wladimir~A.}\
  \bibnamefont {Benalcazar}}, \bibinfo {author} {\bibfnamefont {Taylor~L.}\
  \bibnamefont {Hughes}}, \ and\ \bibinfo {author} {\bibfnamefont {Gaurav}\
  \bibnamefont {Bahl}},\ }\bibfield  {title} {\enquote {\bibinfo {title} {{A
  quantized microwave quadrupole insulator with topologically protected corner
  states}},}\ }\href {\doibase 10.1038/nature25777} {\bibfield  {journal}
  {\bibinfo  {journal} {Nature}\ }\textbf {\bibinfo {volume} {555}},\ \bibinfo
  {pages} {346--350} (\bibinfo {year} {2018})}\BibitemShut {NoStop}%
\bibitem [{\citenamefont {Noh}\ \emph {et~al.}(2018)\citenamefont {Noh},
  \citenamefont {Benalcazar}, \citenamefont {Huang}, \citenamefont {Collins},
  \citenamefont {Chen}, \citenamefont {Hughes},\ and\ \citenamefont
  {Rechtsman}}]{Noh2018}%
  \BibitemOpen
  \bibfield  {author} {\bibinfo {author} {\bibfnamefont {Jiho}\ \bibnamefont
  {Noh}}, \bibinfo {author} {\bibfnamefont {Wladimir~A.}\ \bibnamefont
  {Benalcazar}}, \bibinfo {author} {\bibfnamefont {Sheng}\ \bibnamefont
  {Huang}}, \bibinfo {author} {\bibfnamefont {Matthew~J.}\ \bibnamefont
  {Collins}}, \bibinfo {author} {\bibfnamefont {Kevin~P.}\ \bibnamefont
  {Chen}}, \bibinfo {author} {\bibfnamefont {Taylor~L.}\ \bibnamefont
  {Hughes}}, \ and\ \bibinfo {author} {\bibfnamefont {Mikael~C.}\ \bibnamefont
  {Rechtsman}},\ }\bibfield  {title} {\enquote {\bibinfo {title} {{Topological
  protection of photonic mid-gap defect modes}},}\ }\href {\doibase
  10.1038/s41566-018-0179-3} {\bibfield  {journal} {\bibinfo  {journal} {Nature
  Photonics}\ }\textbf {\bibinfo {volume} {12}},\ \bibinfo {pages} {408--415}
  (\bibinfo {year} {2018})}\BibitemShut {NoStop}%
\bibitem [{\citenamefont {Schindler}\ \emph
  {et~al.}(2018{\natexlab{b}})\citenamefont {Schindler}, \citenamefont {Wang},
  \citenamefont {Vergniory}, \citenamefont {Cook}, \citenamefont {Murani},
  \citenamefont {Sengupta}, \citenamefont {Kasumov}, \citenamefont {Deblock},
  \citenamefont {Jeon}, \citenamefont {Drozdov}, \citenamefont {Bouchiat},
  \citenamefont {Gu{\'{e}}ron}, \citenamefont {Yazdani}, \citenamefont
  {Bernevig},\ and\ \citenamefont {Neupert}}]{Schindler2018-2}%
  \BibitemOpen
  \bibfield  {author} {\bibinfo {author} {\bibfnamefont {Frank}\ \bibnamefont
  {Schindler}}, \bibinfo {author} {\bibfnamefont {Zhijun}\ \bibnamefont
  {Wang}}, \bibinfo {author} {\bibfnamefont {Maia~G.}\ \bibnamefont
  {Vergniory}}, \bibinfo {author} {\bibfnamefont {Ashley~M.}\ \bibnamefont
  {Cook}}, \bibinfo {author} {\bibfnamefont {Anil}\ \bibnamefont {Murani}},
  \bibinfo {author} {\bibfnamefont {Shamashis}\ \bibnamefont {Sengupta}},
  \bibinfo {author} {\bibfnamefont {Alik~Yu.}\ \bibnamefont {Kasumov}},
  \bibinfo {author} {\bibfnamefont {Richard}\ \bibnamefont {Deblock}}, \bibinfo
  {author} {\bibfnamefont {Sangjun}\ \bibnamefont {Jeon}}, \bibinfo {author}
  {\bibfnamefont {Ilya}\ \bibnamefont {Drozdov}}, \bibinfo {author}
  {\bibfnamefont {H{\'{e}}l{\`{e}}ne}\ \bibnamefont {Bouchiat}}, \bibinfo
  {author} {\bibfnamefont {Sophie}\ \bibnamefont {Gu{\'{e}}ron}}, \bibinfo
  {author} {\bibfnamefont {Ali}\ \bibnamefont {Yazdani}}, \bibinfo {author}
  {\bibfnamefont {B.~Andrei}\ \bibnamefont {Bernevig}}, \ and\ \bibinfo
  {author} {\bibfnamefont {Titus}\ \bibnamefont {Neupert}},\ }\bibfield
  {title} {\enquote {\bibinfo {title} {{Higher-order topology in bismuth}},}\
  }\href {\doibase 10.1038/s41567-018-0224-7} {\bibfield  {journal} {\bibinfo
  {journal} {Nature Physics}\ }\textbf {\bibinfo {volume} {14}},\ \bibinfo
  {pages} {918--924} (\bibinfo {year} {2018}{\natexlab{b}})}\BibitemShut
  {NoStop}%
\bibitem [{\citenamefont {Imhof}\ \emph {et~al.}(2018)\citenamefont {Imhof},
  \citenamefont {Berger}, \citenamefont {Bayer}, \citenamefont {Brehm},
  \citenamefont {Molenkamp}, \citenamefont {Kiessling}, \citenamefont
  {Schindler}, \citenamefont {Lee}, \citenamefont {Greiter}, \citenamefont
  {Neupert},\ and\ \citenamefont {Thomale}}]{Imhof2018}%
  \BibitemOpen
  \bibfield  {author} {\bibinfo {author} {\bibfnamefont {Stefan}\ \bibnamefont
  {Imhof}}, \bibinfo {author} {\bibfnamefont {Christian}\ \bibnamefont
  {Berger}}, \bibinfo {author} {\bibfnamefont {Florian}\ \bibnamefont {Bayer}},
  \bibinfo {author} {\bibfnamefont {Johannes}\ \bibnamefont {Brehm}}, \bibinfo
  {author} {\bibfnamefont {Laurens~W.}\ \bibnamefont {Molenkamp}}, \bibinfo
  {author} {\bibfnamefont {Tobias}\ \bibnamefont {Kiessling}}, \bibinfo
  {author} {\bibfnamefont {Frank}\ \bibnamefont {Schindler}}, \bibinfo {author}
  {\bibfnamefont {Ching~Hua}\ \bibnamefont {Lee}}, \bibinfo {author}
  {\bibfnamefont {Martin}\ \bibnamefont {Greiter}}, \bibinfo {author}
  {\bibfnamefont {Titus}\ \bibnamefont {Neupert}}, \ and\ \bibinfo {author}
  {\bibfnamefont {Ronny}\ \bibnamefont {Thomale}},\ }\bibfield  {title}
  {\enquote {\bibinfo {title} {{Topolectrical-circuit realization of
  topological corner modes}},}\ }\href {\doibase 10.1038/s41567-018-0246-1}
  {\bibfield  {journal} {\bibinfo  {journal} {Nature Physics}\ }\textbf
  {\bibinfo {volume} {14}},\ \bibinfo {pages} {925--929} (\bibinfo {year}
  {2018})}\BibitemShut {NoStop}%
\bibitem [{\citenamefont {Xue}\ \emph {et~al.}(2019)\citenamefont {Xue},
  \citenamefont {Yang}, \citenamefont {Gao}, \citenamefont {Chong},\ and\
  \citenamefont {Zhang}}]{xue2019acoustic}%
  \BibitemOpen
  \bibfield  {author} {\bibinfo {author} {\bibfnamefont {Haoran}\ \bibnamefont
  {Xue}}, \bibinfo {author} {\bibfnamefont {Yahui}\ \bibnamefont {Yang}},
  \bibinfo {author} {\bibfnamefont {Fei}\ \bibnamefont {Gao}}, \bibinfo
  {author} {\bibfnamefont {Yidong}\ \bibnamefont {Chong}}, \ and\ \bibinfo
  {author} {\bibfnamefont {Baile}\ \bibnamefont {Zhang}},\ }\bibfield  {title}
  {\enquote {\bibinfo {title} {Acoustic higher-order topological insulator on a
  kagome lattice},}\ }\href {https://www.nature.com/articles/s41563-018-0251-x}
  {\bibfield  {journal} {\bibinfo  {journal} {Nature materials}\ }\textbf
  {\bibinfo {volume} {18}},\ \bibinfo {pages} {108--112} (\bibinfo {year}
  {2019})}\BibitemShut {NoStop}%
\bibitem [{\citenamefont {Ni}\ \emph {et~al.}(2019)\citenamefont {Ni},
  \citenamefont {Weiner}, \citenamefont {Al{\`u}},\ and\ \citenamefont
  {Khanikaev}}]{ni2019observation}%
  \BibitemOpen
  \bibfield  {author} {\bibinfo {author} {\bibfnamefont {Xiang}\ \bibnamefont
  {Ni}}, \bibinfo {author} {\bibfnamefont {Matthew}\ \bibnamefont {Weiner}},
  \bibinfo {author} {\bibfnamefont {Andrea}\ \bibnamefont {Al{\`u}}}, \ and\
  \bibinfo {author} {\bibfnamefont {Alexander~B}\ \bibnamefont {Khanikaev}},\
  }\bibfield  {title} {\enquote {\bibinfo {title} {Observation of higher-order
  topological acoustic states protected by generalized chiral symmetry},}\
  }\href {https://www.nature.com/articles/s41563-018-0252-9} {\bibfield
  {journal} {\bibinfo  {journal} {Nature materials}\ }\textbf {\bibinfo
  {volume} {18}},\ \bibinfo {pages} {113--120} (\bibinfo {year}
  {2019})}\BibitemShut {NoStop}%
\bibitem [{\citenamefont {Li}\ and\ \citenamefont
  {Sun}(2020)}]{PhysRevLett.124.036401}%
  \BibitemOpen
  \bibfield  {author} {\bibinfo {author} {\bibfnamefont {Heqiu}\ \bibnamefont
  {Li}}\ and\ \bibinfo {author} {\bibfnamefont {Kai}\ \bibnamefont {Sun}},\
  }\bibfield  {title} {\enquote {\bibinfo {title} {Pfaffian formalism for
  higher-order topological insulators},}\ }\href {\doibase
  10.1103/PhysRevLett.124.036401} {\bibfield  {journal} {\bibinfo  {journal}
  {Phys. Rev. Lett.}\ }\textbf {\bibinfo {volume} {124}},\ \bibinfo {pages}
  {036401} (\bibinfo {year} {2020})}\BibitemShut {NoStop}%
\bibitem [{\citenamefont {Kudo}\ \emph {et~al.}(2019)\citenamefont {Kudo},
  \citenamefont {Yoshida},\ and\ \citenamefont
  {Hatsugai}}]{PhysRevLett.123.196402}%
  \BibitemOpen
  \bibfield  {author} {\bibinfo {author} {\bibfnamefont {Koji}\ \bibnamefont
  {Kudo}}, \bibinfo {author} {\bibfnamefont {Tsuneya}\ \bibnamefont {Yoshida}},
  \ and\ \bibinfo {author} {\bibfnamefont {Yasuhiro}\ \bibnamefont
  {Hatsugai}},\ }\bibfield  {title} {\enquote {\bibinfo {title} {Higher-order
  topological mott insulators},}\ }\href {\doibase
  10.1103/PhysRevLett.123.196402} {\bibfield  {journal} {\bibinfo  {journal}
  {Phys. Rev. Lett.}\ }\textbf {\bibinfo {volume} {123}},\ \bibinfo {pages}
  {196402} (\bibinfo {year} {2019})}\BibitemShut {NoStop}%
\bibitem [{\citenamefont {Zhang}\ \emph
  {et~al.}(2019{\natexlab{a}})\citenamefont {Zhang}, \citenamefont {Cole},
  \citenamefont {Wu},\ and\ \citenamefont
  {Das~Sarma}}]{PhysRevLett.123.167001}%
  \BibitemOpen
  \bibfield  {author} {\bibinfo {author} {\bibfnamefont {Rui-Xing}\
  \bibnamefont {Zhang}}, \bibinfo {author} {\bibfnamefont {William~S.}\
  \bibnamefont {Cole}}, \bibinfo {author} {\bibfnamefont {Xianxin}\
  \bibnamefont {Wu}}, \ and\ \bibinfo {author} {\bibfnamefont {S.}~\bibnamefont
  {Das~Sarma}},\ }\bibfield  {title} {\enquote {\bibinfo {title} {Higher-order
  topology and nodal topological superconductivity in fe(se,te)
  heterostructures},}\ }\href {\doibase 10.1103/PhysRevLett.123.167001}
  {\bibfield  {journal} {\bibinfo  {journal} {Phys. Rev. Lett.}\ }\textbf
  {\bibinfo {volume} {123}},\ \bibinfo {pages} {167001} (\bibinfo {year}
  {2019}{\natexlab{a}})}\BibitemShut {NoStop}%
\bibitem [{\citenamefont {Yan}(2019{\natexlab{a}})}]{PhysRevLett.123.177001}%
  \BibitemOpen
  \bibfield  {author} {\bibinfo {author} {\bibfnamefont {Zhongbo}\ \bibnamefont
  {Yan}},\ }\bibfield  {title} {\enquote {\bibinfo {title} {Higher-order
  topological odd-parity superconductors},}\ }\href {\doibase
  10.1103/PhysRevLett.123.177001} {\bibfield  {journal} {\bibinfo  {journal}
  {Phys. Rev. Lett.}\ }\textbf {\bibinfo {volume} {123}},\ \bibinfo {pages}
  {177001} (\bibinfo {year} {2019}{\natexlab{a}})}\BibitemShut {NoStop}%
\bibitem [{\citenamefont {Tiwari}\ \emph {et~al.}(2020)\citenamefont {Tiwari},
  \citenamefont {Li}, \citenamefont {Bernevig}, \citenamefont {Neupert},\ and\
  \citenamefont {Parameswaran}}]{PhysRevLett.124.046801}%
  \BibitemOpen
  \bibfield  {author} {\bibinfo {author} {\bibfnamefont {Apoorv}\ \bibnamefont
  {Tiwari}}, \bibinfo {author} {\bibfnamefont {Ming-Hao}\ \bibnamefont {Li}},
  \bibinfo {author} {\bibfnamefont {B.~A.}\ \bibnamefont {Bernevig}}, \bibinfo
  {author} {\bibfnamefont {Titus}\ \bibnamefont {Neupert}}, \ and\ \bibinfo
  {author} {\bibfnamefont {S.~A.}\ \bibnamefont {Parameswaran}},\ }\bibfield
  {title} {\enquote {\bibinfo {title} {Unhinging the surfaces of higher-order
  topological insulators and superconductors},}\ }\href {\doibase
  10.1103/PhysRevLett.124.046801} {\bibfield  {journal} {\bibinfo  {journal}
  {Phys. Rev. Lett.}\ }\textbf {\bibinfo {volume} {124}},\ \bibinfo {pages}
  {046801} (\bibinfo {year} {2020})}\BibitemShut {NoStop}%
\bibitem [{\citenamefont {Zhang}\ \emph
  {et~al.}(2019{\natexlab{b}})\citenamefont {Zhang}, \citenamefont {Hsu},\ and\
  \citenamefont {Sarma}}]{zhang2019higherorder}%
  \BibitemOpen
  \bibfield  {author} {\bibinfo {author} {\bibfnamefont {Rui-Xing}\
  \bibnamefont {Zhang}}, \bibinfo {author} {\bibfnamefont {Yi-Ting}\
  \bibnamefont {Hsu}}, \ and\ \bibinfo {author} {\bibfnamefont {S.~Das}\
  \bibnamefont {Sarma}},\ }\href@noop {} {\enquote {\bibinfo {title}
  {Higher-order topological dirac superconductors},}\ } (\bibinfo {year}
  {2019}{\natexlab{b}}),\ \Eprint {http://arxiv.org/abs/1909.07980}
  {arXiv:1909.07980 [cond-mat.mes-hall]} \BibitemShut {NoStop}%
\bibitem [{\citenamefont {Zhang}\ and\ \citenamefont
  {Trauzettel}(2020)}]{PhysRevResearch.2.012018}%
  \BibitemOpen
  \bibfield  {author} {\bibinfo {author} {\bibfnamefont {Song-Bo}\ \bibnamefont
  {Zhang}}\ and\ \bibinfo {author} {\bibfnamefont {Bj\"orn}\ \bibnamefont
  {Trauzettel}},\ }\bibfield  {title} {\enquote {\bibinfo {title} {Detection of
  second-order topological superconductors by josephson junctions},}\ }\href
  {\doibase 10.1103/PhysRevResearch.2.012018} {\bibfield  {journal} {\bibinfo
  {journal} {Phys. Rev. Research}\ }\textbf {\bibinfo {volume} {2}},\ \bibinfo
  {pages} {012018} (\bibinfo {year} {2020})}\BibitemShut {NoStop}%
\bibitem [{\citenamefont {Dubinkin}\ \emph {et~al.}(2020)\citenamefont
  {Dubinkin}, \citenamefont {May-Mann},\ and\ \citenamefont
  {Hughes}}]{dubinkin2020lieb}%
  \BibitemOpen
  \bibfield  {author} {\bibinfo {author} {\bibfnamefont {Oleg}\ \bibnamefont
  {Dubinkin}}, \bibinfo {author} {\bibfnamefont {Julian}\ \bibnamefont
  {May-Mann}}, \ and\ \bibinfo {author} {\bibfnamefont {Taylor~L.}\
  \bibnamefont {Hughes}},\ }\href@noop {} {\enquote {\bibinfo {title} {Lieb
  schultz mattis-type theorems and other non-perturbative results for strongly
  correlated systems with conserved dipole moments},}\ } (\bibinfo {year}
  {2020}),\ \Eprint {http://arxiv.org/abs/2001.04477} {arXiv:2001.04477
  [cond-mat.str-el]} \BibitemShut {NoStop}%
\bibitem [{\citenamefont {You}\ \emph {et~al.}(2018)\citenamefont {You},
  \citenamefont {Devakul}, \citenamefont {Burnell},\ and\ \citenamefont
  {Neupert}}]{PhysRevB.98.235102}%
  \BibitemOpen
  \bibfield  {author} {\bibinfo {author} {\bibfnamefont {Yizhi}\ \bibnamefont
  {You}}, \bibinfo {author} {\bibfnamefont {Trithep}\ \bibnamefont {Devakul}},
  \bibinfo {author} {\bibfnamefont {F.~J.}\ \bibnamefont {Burnell}}, \ and\
  \bibinfo {author} {\bibfnamefont {Titus}\ \bibnamefont {Neupert}},\
  }\bibfield  {title} {\enquote {\bibinfo {title} {Higher-order
  symmetry-protected topological states for interacting bosons and fermions},}\
  }\href {\doibase 10.1103/PhysRevB.98.235102} {\bibfield  {journal} {\bibinfo
  {journal} {Phys. Rev. B}\ }\textbf {\bibinfo {volume} {98}},\ \bibinfo
  {pages} {235102} (\bibinfo {year} {2018})}\BibitemShut {NoStop}%
\bibitem [{\citenamefont {Dubinkin}\ and\ \citenamefont
  {Hughes}(2019)}]{PhysRevB.99.235132}%
  \BibitemOpen
  \bibfield  {author} {\bibinfo {author} {\bibfnamefont {Oleg}\ \bibnamefont
  {Dubinkin}}\ and\ \bibinfo {author} {\bibfnamefont {Taylor~L.}\ \bibnamefont
  {Hughes}},\ }\bibfield  {title} {\enquote {\bibinfo {title} {Higher-order
  bosonic topological phases in spin models},}\ }\href {\doibase
  10.1103/PhysRevB.99.235132} {\bibfield  {journal} {\bibinfo  {journal} {Phys.
  Rev. B}\ }\textbf {\bibinfo {volume} {99}},\ \bibinfo {pages} {235132}
  (\bibinfo {year} {2019})}\BibitemShut {NoStop}%
\bibitem [{\citenamefont {Park}\ \emph {et~al.}(2019)\citenamefont {Park},
  \citenamefont {Kim}, \citenamefont {Cho},\ and\ \citenamefont
  {Lee}}]{PhysRevLett.123.216803}%
  \BibitemOpen
  \bibfield  {author} {\bibinfo {author} {\bibfnamefont {Moon~Jip}\
  \bibnamefont {Park}}, \bibinfo {author} {\bibfnamefont {Youngkuk}\
  \bibnamefont {Kim}}, \bibinfo {author} {\bibfnamefont {Gil~Young}\
  \bibnamefont {Cho}}, \ and\ \bibinfo {author} {\bibfnamefont {SungBin}\
  \bibnamefont {Lee}},\ }\bibfield  {title} {\enquote {\bibinfo {title}
  {Higher-order topological insulator in twisted bilayer graphene},}\ }\href
  {\doibase 10.1103/PhysRevLett.123.216803} {\bibfield  {journal} {\bibinfo
  {journal} {Phys. Rev. Lett.}\ }\textbf {\bibinfo {volume} {123}},\ \bibinfo
  {pages} {216803} (\bibinfo {year} {2019})}\BibitemShut {NoStop}%
\bibitem [{\citenamefont {Queiroz}\ \emph {et~al.}(2019)\citenamefont
  {Queiroz}, \citenamefont {Fulga}, \citenamefont {Avraham}, \citenamefont
  {Beidenkopf},\ and\ \citenamefont {Cano}}]{PhysRevLett.123.266802}%
  \BibitemOpen
  \bibfield  {author} {\bibinfo {author} {\bibfnamefont {Raquel}\ \bibnamefont
  {Queiroz}}, \bibinfo {author} {\bibfnamefont {Ion~Cosma}\ \bibnamefont
  {Fulga}}, \bibinfo {author} {\bibfnamefont {Nurit}\ \bibnamefont {Avraham}},
  \bibinfo {author} {\bibfnamefont {Haim}\ \bibnamefont {Beidenkopf}}, \ and\
  \bibinfo {author} {\bibfnamefont {Jennifer}\ \bibnamefont {Cano}},\
  }\bibfield  {title} {\enquote {\bibinfo {title} {Partial lattice defects in
  higher-order topological insulators},}\ }\href {\doibase
  10.1103/PhysRevLett.123.266802} {\bibfield  {journal} {\bibinfo  {journal}
  {Phys. Rev. Lett.}\ }\textbf {\bibinfo {volume} {123}},\ \bibinfo {pages}
  {266802} (\bibinfo {year} {2019})}\BibitemShut {NoStop}%
\bibitem [{\citenamefont {Zeng}\ \emph {et~al.}(2020)\citenamefont {Zeng},
  \citenamefont {Yang},\ and\ \citenamefont {Xu}}]{PhysRevB.101.241104}%
  \BibitemOpen
  \bibfield  {author} {\bibinfo {author} {\bibfnamefont {Qi-Bo}\ \bibnamefont
  {Zeng}}, \bibinfo {author} {\bibfnamefont {Yan-Bin}\ \bibnamefont {Yang}}, \
  and\ \bibinfo {author} {\bibfnamefont {Yong}\ \bibnamefont {Xu}},\ }\bibfield
   {title} {\enquote {\bibinfo {title} {Higher-order topological insulators and
  semimetals in generalized aubry-andr\'e-harper models},}\ }\href {\doibase
  10.1103/PhysRevB.101.241104} {\bibfield  {journal} {\bibinfo  {journal}
  {Phys. Rev. B}\ }\textbf {\bibinfo {volume} {101}},\ \bibinfo {pages}
  {241104} (\bibinfo {year} {2020})}\BibitemShut {NoStop}%
\bibitem [{\citenamefont {Kheirkhah}\ \emph
  {et~al.}(2020{\natexlab{a}})\citenamefont {Kheirkhah}, \citenamefont {Nagai},
  \citenamefont {Chen},\ and\ \citenamefont {Marsiglio}}]{kheirkhah1}%
  \BibitemOpen
  \bibfield  {author} {\bibinfo {author} {\bibfnamefont {Majid}\ \bibnamefont
  {Kheirkhah}}, \bibinfo {author} {\bibfnamefont {Yuki}\ \bibnamefont {Nagai}},
  \bibinfo {author} {\bibfnamefont {Chun}\ \bibnamefont {Chen}}, \ and\
  \bibinfo {author} {\bibfnamefont {Frank}\ \bibnamefont {Marsiglio}},\
  }\bibfield  {title} {\enquote {\bibinfo {title} {Majorana corner flat bands
  in two-dimensional second-order topological superconductors},}\ }\href
  {\doibase 10.1103/PhysRevB.101.104502} {\bibfield  {journal} {\bibinfo
  {journal} {Phys. Rev. B}\ }\textbf {\bibinfo {volume} {101}},\ \bibinfo
  {pages} {104502} (\bibinfo {year} {2020}{\natexlab{a}})}\BibitemShut
  {NoStop}%
\bibitem [{\citenamefont {Hsu}\ \emph {et~al.}(2018)\citenamefont {Hsu},
  \citenamefont {Stano}, \citenamefont {Klinovaja},\ and\ \citenamefont
  {Loss}}]{Loss2018}%
  \BibitemOpen
  \bibfield  {author} {\bibinfo {author} {\bibfnamefont {Chen-Hsuan}\
  \bibnamefont {Hsu}}, \bibinfo {author} {\bibfnamefont {Peter}\ \bibnamefont
  {Stano}}, \bibinfo {author} {\bibfnamefont {Jelena}\ \bibnamefont
  {Klinovaja}}, \ and\ \bibinfo {author} {\bibfnamefont {Daniel}\ \bibnamefont
  {Loss}},\ }\bibfield  {title} {\enquote {\bibinfo {title} {Majorana kramers
  pairs in higher-order topological insulators},}\ }\href {\doibase
  10.1103/PhysRevLett.121.196801} {\bibfield  {journal} {\bibinfo  {journal}
  {Phys. Rev. Lett.}\ }\textbf {\bibinfo {volume} {121}},\ \bibinfo {pages}
  {196801} (\bibinfo {year} {2018})}\BibitemShut {NoStop}%
\bibitem [{\citenamefont {Kheirkhah}\ \emph
  {et~al.}(2020{\natexlab{b}})\citenamefont {Kheirkhah}, \citenamefont {Yan},
  \citenamefont {Nagai},\ and\ \citenamefont {Marsiglio}}]{kheirkhah2}%
  \BibitemOpen
  \bibfield  {author} {\bibinfo {author} {\bibfnamefont {Majid}\ \bibnamefont
  {Kheirkhah}}, \bibinfo {author} {\bibfnamefont {Zhongbo}\ \bibnamefont
  {Yan}}, \bibinfo {author} {\bibfnamefont {Yuki}\ \bibnamefont {Nagai}}, \
  and\ \bibinfo {author} {\bibfnamefont {Frank}\ \bibnamefont {Marsiglio}},\
  }\bibfield  {title} {\enquote {\bibinfo {title} {First- and second-order
  topological superconductivity and temperature-driven topological phase
  transitions in the extended hubbard model with spin-orbit coupling},}\ }\href
  {\doibase 10.1103/PhysRevLett.125.017001} {\bibfield  {journal} {\bibinfo
  {journal} {Phys. Rev. Lett.}\ }\textbf {\bibinfo {volume} {125}},\ \bibinfo
  {pages} {017001} (\bibinfo {year} {2020}{\natexlab{b}})}\BibitemShut
  {NoStop}%
\bibitem [{\citenamefont {Yan}\ \emph {et~al.}(2018)\citenamefont {Yan},
  \citenamefont {Song},\ and\ \citenamefont {Wang}}]{ZhongWang2018}%
  \BibitemOpen
  \bibfield  {author} {\bibinfo {author} {\bibfnamefont {Zhongbo}\ \bibnamefont
  {Yan}}, \bibinfo {author} {\bibfnamefont {Fei}\ \bibnamefont {Song}}, \ and\
  \bibinfo {author} {\bibfnamefont {Zhong}\ \bibnamefont {Wang}},\ }\bibfield
  {title} {\enquote {\bibinfo {title} {Majorana corner modes in a
  high-temperature platform},}\ }\href {\doibase
  10.1103/PhysRevLett.121.096803} {\bibfield  {journal} {\bibinfo  {journal}
  {Phys. Rev. Lett.}\ }\textbf {\bibinfo {volume} {121}},\ \bibinfo {pages}
  {096803} (\bibinfo {year} {2018})}\BibitemShut {NoStop}%
\bibitem [{\citenamefont {Wang}\ \emph {et~al.}(2018)\citenamefont {Wang},
  \citenamefont {Liu}, \citenamefont {Lu},\ and\ \citenamefont
  {Zhang}}]{FanZhang2018}%
  \BibitemOpen
  \bibfield  {author} {\bibinfo {author} {\bibfnamefont {Qiyue}\ \bibnamefont
  {Wang}}, \bibinfo {author} {\bibfnamefont {Cheng-Cheng}\ \bibnamefont {Liu}},
  \bibinfo {author} {\bibfnamefont {Yuan-Ming}\ \bibnamefont {Lu}}, \ and\
  \bibinfo {author} {\bibfnamefont {Fan}\ \bibnamefont {Zhang}},\ }\bibfield
  {title} {\enquote {\bibinfo {title} {High-temperature majorana corner
  states},}\ }\href {\doibase 10.1103/PhysRevLett.121.186801} {\bibfield
  {journal} {\bibinfo  {journal} {Phys. Rev. Lett.}\ }\textbf {\bibinfo
  {volume} {121}},\ \bibinfo {pages} {186801} (\bibinfo {year}
  {2018})}\BibitemShut {NoStop}%
\bibitem [{\citenamefont {Yan}(2019{\natexlab{b}})}]{yanPRL2019}%
  \BibitemOpen
  \bibfield  {author} {\bibinfo {author} {\bibfnamefont {Zhongbo}\ \bibnamefont
  {Yan}},\ }\bibfield  {title} {\enquote {\bibinfo {title} {Higher-order
  topological odd-parity superconductors},}\ }\href {\doibase
  10.1103/PhysRevLett.123.177001} {\bibfield  {journal} {\bibinfo  {journal}
  {Phys. Rev. Lett.}\ }\textbf {\bibinfo {volume} {123}},\ \bibinfo {pages}
  {177001} (\bibinfo {year} {2019}{\natexlab{b}})}\BibitemShut {NoStop}%
\bibitem [{\citenamefont {Bultinck}\ \emph {et~al.}(2019)\citenamefont
  {Bultinck}, \citenamefont {Bernevig},\ and\ \citenamefont
  {Zaletel}}]{bultinck2019three}%
  \BibitemOpen
  \bibfield  {author} {\bibinfo {author} {\bibfnamefont {Nick}\ \bibnamefont
  {Bultinck}}, \bibinfo {author} {\bibfnamefont {B.~Andrei}\ \bibnamefont
  {Bernevig}}, \ and\ \bibinfo {author} {\bibfnamefont {Michael~P.}\
  \bibnamefont {Zaletel}},\ }\bibfield  {title} {\enquote {\bibinfo {title}
  {Three-dimensional superconductors with hybrid higher-order topology},}\
  }\href {\doibase 10.1103/PhysRevB.99.125149} {\bibfield  {journal} {\bibinfo
  {journal} {Phys. Rev. B}\ }\textbf {\bibinfo {volume} {99}},\ \bibinfo
  {pages} {125149} (\bibinfo {year} {2019})}\BibitemShut {NoStop}%
\bibitem [{\citenamefont {Kooi}\ \emph {et~al.}(2019)\citenamefont {Kooi},
  \citenamefont {van Miert},\ and\ \citenamefont {Ortix}}]{kooi2019hybrid}%
  \BibitemOpen
  \bibfield  {author} {\bibinfo {author} {\bibfnamefont {Sander~H.}\
  \bibnamefont {Kooi}}, \bibinfo {author} {\bibfnamefont {Guido}\ \bibnamefont
  {van Miert}}, \ and\ \bibinfo {author} {\bibfnamefont {Carmine}\ \bibnamefont
  {Ortix}},\ }\bibfield  {title} {\enquote {\bibinfo {title} {The hybrid-order
  topology of weak topological insulators},}\ }\href@noop {} {\  (\bibinfo
  {year} {2019})},\ \Eprint {http://arxiv.org/abs/1908.00879} {arXiv:1908.00879
  [cond-mat.mes-hall]} \BibitemShut {NoStop}%
\bibitem [{\citenamefont {Ghorashi}\ \emph
  {et~al.}(2019{\natexlab{b}})\citenamefont {Ghorashi}, \citenamefont {Hu},
  \citenamefont {Hughes},\ and\ \citenamefont {Rossi}}]{Ghorashihosc2019}%
  \BibitemOpen
  \bibfield  {author} {\bibinfo {author} {\bibfnamefont {Sayed Ali~Akbar}\
  \bibnamefont {Ghorashi}}, \bibinfo {author} {\bibfnamefont {Xiang}\
  \bibnamefont {Hu}}, \bibinfo {author} {\bibfnamefont {Taylor~L.}\
  \bibnamefont {Hughes}}, \ and\ \bibinfo {author} {\bibfnamefont {Enrico}\
  \bibnamefont {Rossi}},\ }\bibfield  {title} {\enquote {\bibinfo {title}
  {Second-order dirac superconductors and magnetic field induced majorana hinge
  modes},}\ }\href {\doibase 10.1103/PhysRevB.100.020509} {\bibfield  {journal}
  {\bibinfo  {journal} {Phys. Rev. B}\ }\textbf {\bibinfo {volume} {100}},\
  \bibinfo {pages} {020509} (\bibinfo {year} {2019}{\natexlab{b}})}\BibitemShut
  {NoStop}%
\bibitem [{\citenamefont {Lin}\ and\ \citenamefont {Hughes}(2018)}]{Lin2017}%
  \BibitemOpen
  \bibfield  {author} {\bibinfo {author} {\bibfnamefont {Mao}\ \bibnamefont
  {Lin}}\ and\ \bibinfo {author} {\bibfnamefont {Taylor~L.}\ \bibnamefont
  {Hughes}},\ }\bibfield  {title} {\enquote {\bibinfo {title} {Topological
  quadrupolar semimetals},}\ }\href {\doibase 10.1103/PhysRevB.98.241103}
  {\bibfield  {journal} {\bibinfo  {journal} {Phys. Rev. B}\ }\textbf {\bibinfo
  {volume} {98}},\ \bibinfo {pages} {241103} (\bibinfo {year}
  {2018})}\BibitemShut {NoStop}%
\bibitem [{\citenamefont {C\ifmmode \u{a}\else \u{a}\fi{}lug\ifmmode~\u{a}\else
  \u{a}\fi{}ru}\ \emph {et~al.}(2019)\citenamefont {C\ifmmode \u{a}\else
  \u{a}\fi{}lug\ifmmode~\u{a}\else \u{a}\fi{}ru}, \citenamefont {Juri\ifmmode
  \check{c}\else \v{c}\fi{}i\ifmmode~\acute{c}\else \'{c}\fi{}},\ and\
  \citenamefont {Roy}}]{CAlugAru2018}%
  \BibitemOpen
  \bibfield  {author} {\bibinfo {author} {\bibfnamefont {Dumitru}\ \bibnamefont
  {C\ifmmode \u{a}\else \u{a}\fi{}lug\ifmmode~\u{a}\else \u{a}\fi{}ru}},
  \bibinfo {author} {\bibfnamefont {Vladimir}\ \bibnamefont {Juri\ifmmode
  \check{c}\else \v{c}\fi{}i\ifmmode~\acute{c}\else \'{c}\fi{}}}, \ and\
  \bibinfo {author} {\bibfnamefont {Bitan}\ \bibnamefont {Roy}},\ }\bibfield
  {title} {\enquote {\bibinfo {title} {Higher-order topological phases: A
  general principle of construction},}\ }\href {\doibase
  10.1103/PhysRevB.99.041301} {\bibfield  {journal} {\bibinfo  {journal} {Phys.
  Rev. B}\ }\textbf {\bibinfo {volume} {99}},\ \bibinfo {pages} {041301}
  (\bibinfo {year} {2019})}\BibitemShut {NoStop}%
\bibitem [{\citenamefont {Ezawa}(2018)}]{PhysRevLett.120.026801}%
  \BibitemOpen
  \bibfield  {author} {\bibinfo {author} {\bibfnamefont {Motohiko}\
  \bibnamefont {Ezawa}},\ }\bibfield  {title} {\enquote {\bibinfo {title}
  {Higher-order topological insulators and semimetals on the breathing kagome
  and pyrochlore lattices},}\ }\href {\doibase 10.1103/PhysRevLett.120.026801}
  {\bibfield  {journal} {\bibinfo  {journal} {Phys. Rev. Lett.}\ }\textbf
  {\bibinfo {volume} {120}},\ \bibinfo {pages} {026801} (\bibinfo {year}
  {2018})}\BibitemShut {NoStop}%
\bibitem [{\citenamefont {Wieder}\ \emph
  {et~al.}(2020{\natexlab{a}})\citenamefont {Wieder}, \citenamefont {Wang},
  \citenamefont {Cano}, \citenamefont {Dai}, \citenamefont {Schoop},
  \citenamefont {Bradlyn},\ and\ \citenamefont {Bernevig}}]{Wieder2020}%
  \BibitemOpen
  \bibfield  {author} {\bibinfo {author} {\bibfnamefont {Benjamin~J.}\
  \bibnamefont {Wieder}}, \bibinfo {author} {\bibfnamefont {Zhijun}\
  \bibnamefont {Wang}}, \bibinfo {author} {\bibfnamefont {Jennifer}\
  \bibnamefont {Cano}}, \bibinfo {author} {\bibfnamefont {Xi}~\bibnamefont
  {Dai}}, \bibinfo {author} {\bibfnamefont {Leslie~M.}\ \bibnamefont {Schoop}},
  \bibinfo {author} {\bibfnamefont {Barry}\ \bibnamefont {Bradlyn}}, \ and\
  \bibinfo {author} {\bibfnamefont {B.~Andrei}\ \bibnamefont {Bernevig}},\
  }\bibfield  {title} {\enquote {\bibinfo {title} {Strong and fragile
  topological dirac semimetals with higher-order fermi arcs},}\ }\href
  {\doibase https://doi.org/10.1038/s41467-020-14443-5} {\bibfield  {journal}
  {\bibinfo  {journal} {Nature Communications}\ }\textbf {\bibinfo {volume}
  {11}} (\bibinfo {year} {2020}{\natexlab{a}}),\
  https://doi.org/10.1038/s41467-020-14443-5}\BibitemShut {NoStop}%
\bibitem [{\citenamefont {Ezawa}(2019)}]{ezawa2019second}%
  \BibitemOpen
  \bibfield  {author} {\bibinfo {author} {\bibfnamefont {Motohiko}\
  \bibnamefont {Ezawa}},\ }\bibfield  {title} {\enquote {\bibinfo {title}
  {Second-order topological insulators and loop-nodal semimetals in transition
  metal dichalcogenides xte 2 (x= mo, w)},}\ }\href
  {https://www.nature.com/articles/s41598-019-41746-5} {\bibfield  {journal}
  {\bibinfo  {journal} {Scientific reports}\ }\textbf {\bibinfo {volume} {9}},\
  \bibinfo {pages} {1--11} (\bibinfo {year} {2019})}\BibitemShut {NoStop}%
\bibitem [{\citenamefont {Szabo}\ and\ \citenamefont
  {Roy}(2020)}]{szabo2020dirty}%
  \BibitemOpen
  \bibfield  {author} {\bibinfo {author} {\bibfnamefont {Andras}\ \bibnamefont
  {Szabo}}\ and\ \bibinfo {author} {\bibfnamefont {Bitan}\ \bibnamefont
  {Roy}},\ }\href@noop {} {\enquote {\bibinfo {title} {Dirty higher-order dirac
  semimetal: Quantum criticality and bulk-boundary correspondence},}\ }
  (\bibinfo {year} {2020}),\ \Eprint {http://arxiv.org/abs/2002.09475}
  {arXiv:2002.09475 [cond-mat.mes-hall]} \BibitemShut {NoStop}%
\bibitem [{\citenamefont {Wang}\ \emph {et~al.}(2019)\citenamefont {Wang},
  \citenamefont {Wieder}, \citenamefont {Li}, \citenamefont {Yan},\ and\
  \citenamefont {Bernevig}}]{PhysRevLett.123.186401}%
  \BibitemOpen
  \bibfield  {author} {\bibinfo {author} {\bibfnamefont {Zhijun}\ \bibnamefont
  {Wang}}, \bibinfo {author} {\bibfnamefont {Benjamin~J.}\ \bibnamefont
  {Wieder}}, \bibinfo {author} {\bibfnamefont {Jian}\ \bibnamefont {Li}},
  \bibinfo {author} {\bibfnamefont {Binghai}\ \bibnamefont {Yan}}, \ and\
  \bibinfo {author} {\bibfnamefont {B.~Andrei}\ \bibnamefont {Bernevig}},\
  }\bibfield  {title} {\enquote {\bibinfo {title} {Higher-order topology,
  monopole nodal lines, and the origin of large fermi arcs in transition metal
  dichalcogenides $x{\mathrm{te}}_{2}$ ($x=\mathrm{Mo},\mathrm{W}$)},}\ }\href
  {\doibase 10.1103/PhysRevLett.123.186401} {\bibfield  {journal} {\bibinfo
  {journal} {Phys. Rev. Lett.}\ }\textbf {\bibinfo {volume} {123}},\ \bibinfo
  {pages} {186401} (\bibinfo {year} {2019})}\BibitemShut {NoStop}%
\bibitem [{\citenamefont {Li}\ \emph {et~al.}(2020{\natexlab{b}})\citenamefont
  {Li}, \citenamefont {Wang}, \citenamefont {Li}, \citenamefont {Zheng},
  \citenamefont {Brinkman}, \citenamefont {Yu},\ and\ \citenamefont
  {Liao}}]{PhysRevLett.124.156601}%
  \BibitemOpen
  \bibfield  {author} {\bibinfo {author} {\bibfnamefont {Cai-Zhen}\
  \bibnamefont {Li}}, \bibinfo {author} {\bibfnamefont {An-Qi}\ \bibnamefont
  {Wang}}, \bibinfo {author} {\bibfnamefont {Chuan}\ \bibnamefont {Li}},
  \bibinfo {author} {\bibfnamefont {Wen-Zhuang}\ \bibnamefont {Zheng}},
  \bibinfo {author} {\bibfnamefont {Alexander}\ \bibnamefont {Brinkman}},
  \bibinfo {author} {\bibfnamefont {Da-Peng}\ \bibnamefont {Yu}}, \ and\
  \bibinfo {author} {\bibfnamefont {Zhi-Min}\ \bibnamefont {Liao}},\ }\bibfield
   {title} {\enquote {\bibinfo {title} {Reducing electronic transport dimension
  to topological hinge states by increasing geometry size of dirac semimetal
  josephson junctions},}\ }\href {\doibase 10.1103/PhysRevLett.124.156601}
  {\bibfield  {journal} {\bibinfo  {journal} {Phys. Rev. Lett.}\ }\textbf
  {\bibinfo {volume} {124}},\ \bibinfo {pages} {156601} (\bibinfo {year}
  {2020}{\natexlab{b}})}\BibitemShut {NoStop}%
\bibitem [{\citenamefont {Armitage}\ \emph {et~al.}(2018)\citenamefont
  {Armitage}, \citenamefont {Mele},\ and\ \citenamefont
  {Vishwanath}}]{reviewweyl}%
  \BibitemOpen
  \bibfield  {author} {\bibinfo {author} {\bibfnamefont {N.~P.}\ \bibnamefont
  {Armitage}}, \bibinfo {author} {\bibfnamefont {E.~J.}\ \bibnamefont {Mele}},
  \ and\ \bibinfo {author} {\bibfnamefont {Ashvin}\ \bibnamefont
  {Vishwanath}},\ }\bibfield  {title} {\enquote {\bibinfo {title} {Weyl and
  dirac semimetals in three-dimensional solids},}\ }\href {\doibase
  10.1103/RevModPhys.90.015001} {\bibfield  {journal} {\bibinfo  {journal}
  {Rev. Mod. Phys.}\ }\textbf {\bibinfo {volume} {90}},\ \bibinfo {pages}
  {015001} (\bibinfo {year} {2018})}\BibitemShut {NoStop}%
\bibitem [{\citenamefont {Li}\ \emph {et~al.}(2016)\citenamefont {Li},
  \citenamefont {Luo}, \citenamefont {Dai}, \citenamefont {Yu}, \citenamefont
  {Zhang},\ and\ \citenamefont {Chen}}]{hybridWeyl}%
  \BibitemOpen
  \bibfield  {author} {\bibinfo {author} {\bibfnamefont {Fei-Ye}\ \bibnamefont
  {Li}}, \bibinfo {author} {\bibfnamefont {Xi}~\bibnamefont {Luo}}, \bibinfo
  {author} {\bibfnamefont {Xi}~\bibnamefont {Dai}}, \bibinfo {author}
  {\bibfnamefont {Yue}\ \bibnamefont {Yu}}, \bibinfo {author} {\bibfnamefont
  {Fan}\ \bibnamefont {Zhang}}, \ and\ \bibinfo {author} {\bibfnamefont {Gang}\
  \bibnamefont {Chen}},\ }\bibfield  {title} {\enquote {\bibinfo {title}
  {Hybrid weyl semimetal},}\ }\href {\doibase 10.1103/PhysRevB.94.121105}
  {\bibfield  {journal} {\bibinfo  {journal} {Phys. Rev. B}\ }\textbf {\bibinfo
  {volume} {94}},\ \bibinfo {pages} {121105} (\bibinfo {year}
  {2016})}\BibitemShut {NoStop}%
\bibitem [{\citenamefont {Fang}\ \emph {et~al.}(2012)\citenamefont {Fang},
  \citenamefont {Gilbert},\ and\ \citenamefont {Bernevig}}]{fang2012bulk}%
  \BibitemOpen
  \bibfield  {author} {\bibinfo {author} {\bibfnamefont {Chen}\ \bibnamefont
  {Fang}}, \bibinfo {author} {\bibfnamefont {Matthew~J.}\ \bibnamefont
  {Gilbert}}, \ and\ \bibinfo {author} {\bibfnamefont {B.~Andrei}\ \bibnamefont
  {Bernevig}},\ }\bibfield  {title} {\enquote {\bibinfo {title} {Bulk
  topological invariants in noninteracting point group symmetric insulators},}\
  }\href {\doibase 10.1103/PhysRevB.86.115112} {\bibfield  {journal} {\bibinfo
  {journal} {Phys. Rev. B}\ }\textbf {\bibinfo {volume} {86}},\ \bibinfo
  {pages} {115112} (\bibinfo {year} {2012})}\BibitemShut {NoStop}%
\bibitem [{\citenamefont {Benalcazar}\ \emph {et~al.}(2014)\citenamefont
  {Benalcazar}, \citenamefont {Teo},\ and\ \citenamefont
  {Hughes}}]{Classification2014Wladimir}%
  \BibitemOpen
  \bibfield  {author} {\bibinfo {author} {\bibfnamefont {Wladimir~A.}\
  \bibnamefont {Benalcazar}}, \bibinfo {author} {\bibfnamefont {Jeffrey C.~Y.}\
  \bibnamefont {Teo}}, \ and\ \bibinfo {author} {\bibfnamefont {Taylor~L.}\
  \bibnamefont {Hughes}},\ }\bibfield  {title} {\enquote {\bibinfo {title}
  {Classification of two-dimensional topological crystalline superconductors
  and majorana bound states at disclinations},}\ }\href {\doibase
  10.1103/PhysRevB.89.224503} {\bibfield  {journal} {\bibinfo  {journal} {Phys.
  Rev. B}\ }\textbf {\bibinfo {volume} {89}},\ \bibinfo {pages} {224503}
  (\bibinfo {year} {2014})}\BibitemShut {NoStop}%
\bibitem [{sm()}]{sm}%
  \BibitemOpen
  \href@noop {} {\bibinfo  {journal} {Supplementary material}\ }\BibitemShut
  {NoStop}%
\bibitem [{\citenamefont {Peterson}\ \emph
  {et~al.}(2020{\natexlab{a}})\citenamefont {Peterson}, \citenamefont {Li},
  \citenamefont {Benalcazar}, \citenamefont {Hughes},\ and\ \citenamefont
  {Bahl}}]{Peterson1114}%
  \BibitemOpen
\bibfield  {journal} {  }\bibfield  {author} {\bibinfo {author} {\bibfnamefont
  {Christopher~W.}\ \bibnamefont {Peterson}}, \bibinfo {author} {\bibfnamefont
  {Tianhe}\ \bibnamefont {Li}}, \bibinfo {author} {\bibfnamefont {Wladimir~A.}\
  \bibnamefont {Benalcazar}}, \bibinfo {author} {\bibfnamefont {Taylor~L.}\
  \bibnamefont {Hughes}}, \ and\ \bibinfo {author} {\bibfnamefont {Gaurav}\
  \bibnamefont {Bahl}},\ }\bibfield  {title} {\enquote {\bibinfo {title} {A
  fractional corner anomaly reveals higher-order topology},}\ }\href {\doibase
  10.1126/science.aba7604} {\bibfield  {journal} {\bibinfo  {journal}
  {Science}\ }\textbf {\bibinfo {volume} {368}},\ \bibinfo {pages} {1114--1118}
  (\bibinfo {year} {2020}{\natexlab{a}})}\BibitemShut {NoStop}%
\bibitem [{\citenamefont {Peterson}\ \emph
  {et~al.}(2020{\natexlab{b}})\citenamefont {Peterson}, \citenamefont {Li},
  \citenamefont {Jiang}, \citenamefont {Hughes},\ and\ \citenamefont
  {Bahl}}]{peterson2020observation}%
  \BibitemOpen
  \bibfield  {author} {\bibinfo {author} {\bibfnamefont {Christopher~W.}\
  \bibnamefont {Peterson}}, \bibinfo {author} {\bibfnamefont {Tianhe}\
  \bibnamefont {Li}}, \bibinfo {author} {\bibfnamefont {Wentao}\ \bibnamefont
  {Jiang}}, \bibinfo {author} {\bibfnamefont {Taylor~L.}\ \bibnamefont
  {Hughes}}, \ and\ \bibinfo {author} {\bibfnamefont {Gaurav}\ \bibnamefont
  {Bahl}},\ }\bibfield  {title} {\enquote {\bibinfo {title} {Observation of
  trapped fractional charge and topological states at disclination defects in
  higher-order topological insulators},}\ }\href@noop {} {\  (\bibinfo {year}
  {2020}{\natexlab{b}})},\ \Eprint {http://arxiv.org/abs/2004.11390}
  {arXiv:2004.11390 [cond-mat.mes-hall]} \BibitemShut {NoStop}%
\bibitem [{\citenamefont {Geier}\ \emph {et~al.}(2018)\citenamefont {Geier},
  \citenamefont {Trifunovic}, \citenamefont {Hoskam},\ and\ \citenamefont
  {Brouwer}}]{extrinsichoti1}%
  \BibitemOpen
  \bibfield  {author} {\bibinfo {author} {\bibfnamefont {Max}\ \bibnamefont
  {Geier}}, \bibinfo {author} {\bibfnamefont {Luka}\ \bibnamefont
  {Trifunovic}}, \bibinfo {author} {\bibfnamefont {Max}\ \bibnamefont
  {Hoskam}}, \ and\ \bibinfo {author} {\bibfnamefont {Piet~W.}\ \bibnamefont
  {Brouwer}},\ }\bibfield  {title} {\enquote {\bibinfo {title} {Second-order
  topological insulators and superconductors with an order-two crystalline
  symmetry},}\ }\href {\doibase 10.1103/PhysRevB.97.205135} {\bibfield
  {journal} {\bibinfo  {journal} {Phys. Rev. B}\ }\textbf {\bibinfo {volume}
  {97}},\ \bibinfo {pages} {205135} (\bibinfo {year} {2018})}\BibitemShut
  {NoStop}%
\bibitem [{\citenamefont {Yang}\ \emph {et~al.}(2011)\citenamefont {Yang},
  \citenamefont {Lu},\ and\ \citenamefont {Ran}}]{cdw1}%
  \BibitemOpen
  \bibfield  {author} {\bibinfo {author} {\bibfnamefont {Kai-Yu}\ \bibnamefont
  {Yang}}, \bibinfo {author} {\bibfnamefont {Yuan-Ming}\ \bibnamefont {Lu}}, \
  and\ \bibinfo {author} {\bibfnamefont {Ying}\ \bibnamefont {Ran}},\
  }\bibfield  {title} {\enquote {\bibinfo {title} {Quantum hall effects in a
  weyl semimetal: Possible application in pyrochlore iridates},}\ }\href
  {\doibase 10.1103/PhysRevB.84.075129} {\bibfield  {journal} {\bibinfo
  {journal} {Phys. Rev. B}\ }\textbf {\bibinfo {volume} {84}},\ \bibinfo
  {pages} {075129} (\bibinfo {year} {2011})}\BibitemShut {NoStop}%
\bibitem [{\citenamefont {Wang}\ and\ \citenamefont {Zhang}(2013)}]{cdw2}%
  \BibitemOpen
  \bibfield  {author} {\bibinfo {author} {\bibfnamefont {Zhong}\ \bibnamefont
  {Wang}}\ and\ \bibinfo {author} {\bibfnamefont {Shou-Cheng}\ \bibnamefont
  {Zhang}},\ }\bibfield  {title} {\enquote {\bibinfo {title} {Chiral anomaly,
  charge density waves, and axion strings from weyl semimetals},}\ }\href
  {\doibase 10.1103/PhysRevB.87.161107} {\bibfield  {journal} {\bibinfo
  {journal} {Phys. Rev. B}\ }\textbf {\bibinfo {volume} {87}},\ \bibinfo
  {pages} {161107} (\bibinfo {year} {2013})}\BibitemShut {NoStop}%
\bibitem [{\citenamefont {You}\ \emph {et~al.}(2016)\citenamefont {You},
  \citenamefont {Cho},\ and\ \citenamefont {Hughes}}]{cdw3}%
  \BibitemOpen
  \bibfield  {author} {\bibinfo {author} {\bibfnamefont {Yizhi}\ \bibnamefont
  {You}}, \bibinfo {author} {\bibfnamefont {Gil~Young}\ \bibnamefont {Cho}}, \
  and\ \bibinfo {author} {\bibfnamefont {Taylor~L.}\ \bibnamefont {Hughes}},\
  }\bibfield  {title} {\enquote {\bibinfo {title} {Response properties of axion
  insulators and weyl semimetals driven by screw dislocations and dynamical
  axion strings},}\ }\href {\doibase 10.1103/PhysRevB.94.085102} {\bibfield
  {journal} {\bibinfo  {journal} {Phys. Rev. B}\ }\textbf {\bibinfo {volume}
  {94}},\ \bibinfo {pages} {085102} (\bibinfo {year} {2016})}\BibitemShut
  {NoStop}%
\bibitem [{\citenamefont {Gooth}\ \emph {et~al.}(2019)\citenamefont {Gooth},
  \citenamefont {Bradlyn}, \citenamefont {Honnali}, \citenamefont {Schindler},
  \citenamefont {Kumar}, \citenamefont {Noky}, \citenamefont {Qi},
  \citenamefont {Shekhar}, \citenamefont {Sun}, \citenamefont {Wang} \emph
  {et~al.}}]{cdw4}%
  \BibitemOpen
  \bibfield  {author} {\bibinfo {author} {\bibfnamefont {J}~\bibnamefont
  {Gooth}}, \bibinfo {author} {\bibfnamefont {B}~\bibnamefont {Bradlyn}},
  \bibinfo {author} {\bibfnamefont {S}~\bibnamefont {Honnali}}, \bibinfo
  {author} {\bibfnamefont {C}~\bibnamefont {Schindler}}, \bibinfo {author}
  {\bibfnamefont {N}~\bibnamefont {Kumar}}, \bibinfo {author} {\bibfnamefont
  {J}~\bibnamefont {Noky}}, \bibinfo {author} {\bibfnamefont {Y}~\bibnamefont
  {Qi}}, \bibinfo {author} {\bibfnamefont {C}~\bibnamefont {Shekhar}}, \bibinfo
  {author} {\bibfnamefont {Y}~\bibnamefont {Sun}}, \bibinfo {author}
  {\bibfnamefont {Z}~\bibnamefont {Wang}},  \emph {et~al.},\ }\bibfield
  {title} {\enquote {\bibinfo {title} {Axionic charge-density wave in the weyl
  semimetal (tase 4) 2 i},}\ }\href
  {https://www.nature.com/articles/s41586-019-1630-4} {\bibfield  {journal}
  {\bibinfo  {journal} {Nature}\ }\textbf {\bibinfo {volume} {575}},\ \bibinfo
  {pages} {315--319} (\bibinfo {year} {2019})}\BibitemShut {NoStop}%
\bibitem [{\citenamefont {Wei}\ \emph {et~al.}(2012)\citenamefont {Wei},
  \citenamefont {Chao},\ and\ \citenamefont {Aji}}]{cdw5}%
  \BibitemOpen
  \bibfield  {author} {\bibinfo {author} {\bibfnamefont {Huazhou}\ \bibnamefont
  {Wei}}, \bibinfo {author} {\bibfnamefont {Sung-Po}\ \bibnamefont {Chao}}, \
  and\ \bibinfo {author} {\bibfnamefont {Vivek}\ \bibnamefont {Aji}},\
  }\bibfield  {title} {\enquote {\bibinfo {title} {Excitonic phases from weyl
  semimetals},}\ }\href {\doibase 10.1103/PhysRevLett.109.196403} {\bibfield
  {journal} {\bibinfo  {journal} {Phys. Rev. Lett.}\ }\textbf {\bibinfo
  {volume} {109}},\ \bibinfo {pages} {196403} (\bibinfo {year}
  {2012})}\BibitemShut {NoStop}%
\bibitem [{\citenamefont {Wieder}\ \emph
  {et~al.}(2020{\natexlab{b}})\citenamefont {Wieder}, \citenamefont {Lin},\
  and\ \citenamefont {Bradlyn}}]{wieder2020dynamical}%
  \BibitemOpen
  \bibfield  {author} {\bibinfo {author} {\bibfnamefont {Benjamin~J.}\
  \bibnamefont {Wieder}}, \bibinfo {author} {\bibfnamefont {Kuan-Sen}\
  \bibnamefont {Lin}}, \ and\ \bibinfo {author} {\bibfnamefont {Barry}\
  \bibnamefont {Bradlyn}},\ }\bibfield  {title} {\enquote {\bibinfo {title} {Is
  the dynamical axion weyl-charge-density wave an axionic band insulator?}}\
  }\href@noop {} {\  (\bibinfo {year} {2020}{\natexlab{b}})},\ \Eprint
  {http://arxiv.org/abs/2004.11401} {arXiv:2004.11401 [cond-mat.mes-hall]}
  \BibitemShut {NoStop}%
\bibitem [{\citenamefont {Vazifeh}\ and\ \citenamefont
  {Franz}(2013)}]{vazifeh}%
  \BibitemOpen
  \bibfield  {author} {\bibinfo {author} {\bibfnamefont {M.~M.}\ \bibnamefont
  {Vazifeh}}\ and\ \bibinfo {author} {\bibfnamefont {M.}~\bibnamefont
  {Franz}},\ }\bibfield  {title} {\enquote {\bibinfo {title} {Electromagnetic
  response of weyl semimetals},}\ }\href {\doibase
  10.1103/PhysRevLett.111.027201} {\bibfield  {journal} {\bibinfo  {journal}
  {Phys. Rev. Lett.}\ }\textbf {\bibinfo {volume} {111}},\ \bibinfo {pages}
  {027201} (\bibinfo {year} {2013})}\BibitemShut {NoStop}%
\bibitem [{\citenamefont {Burkov}\ and\ \citenamefont
  {Balents}(2011)}]{burkovPRL2011}%
  \BibitemOpen
  \bibfield  {author} {\bibinfo {author} {\bibfnamefont {A.~A.}\ \bibnamefont
  {Burkov}}\ and\ \bibinfo {author} {\bibfnamefont {Leon}\ \bibnamefont
  {Balents}},\ }\bibfield  {title} {\enquote {\bibinfo {title} {Weyl semimetal
  in a topological insulator multilayer},}\ }\href {\doibase
  10.1103/PhysRevLett.107.127205} {\bibfield  {journal} {\bibinfo  {journal}
  {Phys. Rev. Lett.}\ }\textbf {\bibinfo {volume} {107}},\ \bibinfo {pages}
  {127205} (\bibinfo {year} {2011})}\BibitemShut {NoStop}%
\bibitem [{\citenamefont {Wan}\ \emph {et~al.}(2011)\citenamefont {Wan},
  \citenamefont {Turner}, \citenamefont {Vishwanath},\ and\ \citenamefont
  {Savrasov}}]{PhysRevB.83.205101}%
  \BibitemOpen
  \bibfield  {author} {\bibinfo {author} {\bibfnamefont {Xiangang}\
  \bibnamefont {Wan}}, \bibinfo {author} {\bibfnamefont {Ari~M.}\ \bibnamefont
  {Turner}}, \bibinfo {author} {\bibfnamefont {Ashvin}\ \bibnamefont
  {Vishwanath}}, \ and\ \bibinfo {author} {\bibfnamefont {Sergey~Y.}\
  \bibnamefont {Savrasov}},\ }\bibfield  {title} {\enquote {\bibinfo {title}
  {Topological semimetal and fermi-arc surface states in the electronic
  structure of pyrochlore iridates},}\ }\href {\doibase
  10.1103/PhysRevB.83.205101} {\bibfield  {journal} {\bibinfo  {journal} {Phys.
  Rev. B}\ }\textbf {\bibinfo {volume} {83}},\ \bibinfo {pages} {205101}
  (\bibinfo {year} {2011})}\BibitemShut {NoStop}%
\bibitem [{\citenamefont {Soluyanov}\ \emph {et~al.}(2015)\citenamefont
  {Soluyanov}, \citenamefont {Gresch}, \citenamefont {Wang}, \citenamefont
  {Wu}, \citenamefont {Troyer}, \citenamefont {Dai},\ and\ \citenamefont
  {Bernevig}}]{Soluyanov2015}%
  \BibitemOpen
  \bibfield  {author} {\bibinfo {author} {\bibfnamefont {Alexey~A.}\
  \bibnamefont {Soluyanov}}, \bibinfo {author} {\bibfnamefont {Dominik}\
  \bibnamefont {Gresch}}, \bibinfo {author} {\bibfnamefont {Zhijun}\
  \bibnamefont {Wang}}, \bibinfo {author} {\bibfnamefont {QuanSheng}\
  \bibnamefont {Wu}}, \bibinfo {author} {\bibfnamefont {Matthias}\ \bibnamefont
  {Troyer}}, \bibinfo {author} {\bibfnamefont {Xi}~\bibnamefont {Dai}}, \ and\
  \bibinfo {author} {\bibfnamefont {B.~Andrei}\ \bibnamefont {Bernevig}},\
  }\bibfield  {title} {\enquote {\bibinfo {title} {Type-ii weyl semimetals},}\
  }\href {\doibase 10.1038/nature15768} {\bibfield  {journal} {\bibinfo
  {journal} {Nature}\ }\textbf {\bibinfo {volume} {527}},\ \bibinfo {pages}
  {495--498} (\bibinfo {year} {2015})}\BibitemShut {NoStop}%
\bibitem [{\citenamefont {Ghorashi}(2020)}]{GhorashiFloquet2}%
  \BibitemOpen
  \bibfield  {author} {\bibinfo {author} {\bibfnamefont {Sayed Ali~Akbar}\
  \bibnamefont {Ghorashi}},\ }\bibfield  {title} {\enquote {\bibinfo {title}
  {Hybrid dispersion dirac semimetal and hybrid weyl phases in luttinger
  semimetals: A dynamical approach},}\ }\href {\doibase 10.1002/andp.201900336}
  {\bibfield  {journal} {\bibinfo  {journal} {Annalen der Physik}\ }\textbf
  {\bibinfo {volume} {532}},\ \bibinfo {pages} {1900336} (\bibinfo {year}
  {2020})}\BibitemShut {NoStop}%
\bibitem [{\citenamefont {Hubener}\ \emph {et~al.}(2017)\citenamefont
  {Hubener}, \citenamefont {Sentef}, \citenamefont {De~Giovannini},
  \citenamefont {Kemper},\ and\ \citenamefont {Rubio}}]{hubener2017creating}%
  \BibitemOpen
  \bibfield  {author} {\bibinfo {author} {\bibfnamefont {Hannes}\ \bibnamefont
  {Hubener}}, \bibinfo {author} {\bibfnamefont {Michael~A}\ \bibnamefont
  {Sentef}}, \bibinfo {author} {\bibfnamefont {Umberto}\ \bibnamefont
  {De~Giovannini}}, \bibinfo {author} {\bibfnamefont {Alexander~F}\
  \bibnamefont {Kemper}}, \ and\ \bibinfo {author} {\bibfnamefont {Angel}\
  \bibnamefont {Rubio}},\ }\bibfield  {title} {\enquote {\bibinfo {title}
  {Creating stable floquet--weyl semimetals by laser-driving of 3d dirac
  materials},}\ }\href {https://www.nature.com/articles/ncomms13940} {\bibfield
   {journal} {\bibinfo  {journal} {Nature communications}\ }\textbf {\bibinfo
  {volume} {8}},\ \bibinfo {pages} {1--8} (\bibinfo {year} {2017})}\BibitemShut
  {NoStop}%
\bibitem [{\citenamefont {Ghorashi}\ \emph {et~al.}(2018)\citenamefont
  {Ghorashi}, \citenamefont {Hosur},\ and\ \citenamefont
  {Ting}}]{GhorashiFloquet1}%
  \BibitemOpen
  \bibfield  {author} {\bibinfo {author} {\bibfnamefont {Sayed Ali~Akbar}\
  \bibnamefont {Ghorashi}}, \bibinfo {author} {\bibfnamefont {Pavan}\
  \bibnamefont {Hosur}}, \ and\ \bibinfo {author} {\bibfnamefont {Chin-Sen}\
  \bibnamefont {Ting}},\ }\bibfield  {title} {\enquote {\bibinfo {title}
  {Irradiated three-dimensional luttinger semimetal: A factory for engineering
  weyl semimetals},}\ }\href {\doibase 10.1103/PhysRevB.97.205402} {\bibfield
  {journal} {\bibinfo  {journal} {Phys. Rev. B}\ }\textbf {\bibinfo {volume}
  {97}},\ \bibinfo {pages} {205402} (\bibinfo {year} {2018})}\BibitemShut
  {NoStop}%
\bibitem [{\citenamefont {Oka}\ and\ \citenamefont
  {Kitamura}(2019)}]{oka2019floquet}%
  \BibitemOpen
  \bibfield  {author} {\bibinfo {author} {\bibfnamefont {Takashi}\ \bibnamefont
  {Oka}}\ and\ \bibinfo {author} {\bibfnamefont {Sota}\ \bibnamefont
  {Kitamura}},\ }\bibfield  {title} {\enquote {\bibinfo {title} {Floquet
  engineering of quantum materials},}\ }\href {\doibase
  10.1146/annurev-conmatphys-031218-013423} {\bibfield  {journal} {\bibinfo
  {journal} {Annual Review of Condensed Matter Physics}\ }\textbf {\bibinfo
  {volume} {10}},\ \bibinfo {pages} {387--408} (\bibinfo {year}
  {2019})}\BibitemShut {NoStop}%
\end{thebibliography}%


%

%

\end{document}